\documentclass[11pt]{article}
\usepackage{mathtools}
\usepackage{amstext,amsmath,amssymb,amsfonts,bbm}
\numberwithin{equation}{section}
\usepackage{hyperref}
\usepackage{authblk}
\usepackage{caption}
\usepackage{subcaption}
\usepackage{tikz-feynman}
\usepackage[utf8]{inputenc}
\usepackage{epsfig}
\usepackage{amsthm}
\usepackage{mathtools}
\usepackage{xcolor}
\usepackage{graphicx}
\usepackage{braket}
\usepackage{slashed}
\usepackage{float}
\usepackage{enumitem}
\usepackage{physics}
\usepackage[lmargin=80pt,rmargin=80pt,tmargin=80pt,bmargin=80pt]{geometry}
\usepackage[normalem]{ulem}
\usepackage{comment}
\usepackage{esint}
\usepackage[normalem]{ulem}
\usepackage{cite}
\usepackage{booktabs}  
\usepackage{multirow,makecell} 
\usepackage[T1]{fontenc}
\usepackage[utf8]{inputenc}\usepackage{tabularx}
\usepackage{makecell}
\usepackage{multirow}

\newcolumntype{Y}{>{\centering\arraybackslash}X}
\newcommand{\be}{\begin{equation}}
\newcommand{\ee}{\end{equation}}

\newcommand{\p}{\partial}

\theoremstyle{remark}

\newcommand{\Z}{{\mathbb Z}}

\newcommand{\T}{{\mathbb T}}
\newcommand{\LP}{{\text{1L}}}
\let\a=\alpha \let\b=\beta  \let\g=\gamma  \let\d=\delta
\let\z=\zeta     \let\th=\theta  \let\k=\kappa \let\l=\lambda
               
\let\s=\sigma \let\t=\tau   \let\vph=\varphi  
    
\let\G=\Gamma \let\D=\Delta   \let\L=\Lambda \let\X=F

\def\NN{{\cal N}} 
  \def\OO{{\cal O}}
\def\DD{{\cal D}}

\newcommand{\gb}{\bar{g}}

\newcommand{\mb}{\bar{m}}

\newcommand{\Gh}{\hat{\Gamma}}

\newcommand{\beq}{\begin{equation}}
\newcommand{\eeq}{\end{equation}}
\newcommand{\bea}{\begin{eqnarray}}
\newcommand{\eea}{\end{eqnarray}}

\begin{document}

\title{\bf Critical Phenomena on the Bethe Lattice}

\author[1,2]{Rudrajit Banerjee}
\author[1]{Nicolas Delporte}
\author[1]{Saswato Sen}
\author[1]{Reiko Toriumi}

\affil[1]{\normalsize\it Okinawa Institute of Science and Technology Graduate University, 1919-1, Tancha, Onna, Kunigami District, Okinawa 904-0495, Japan}
\affil[2]{\normalsize\it Centenary College of Louisiana, 2911 Centenary Boulevard, Shreveport, LA 71104, USA \authorcr
Emails: \url{nicolas.delporte@oist.jp}, \url{rudrajit.banerjee@oist.jp}, \url{rbanerjee@centenary.edu }, \url{saswato.sen@oist.jp}, \url{reiko.toriumi@oist.jp}
}

\date{\vspace{-5ex}}

\maketitle

\hrule\bigskip

\begin{abstract}

\noindent 
We investigate the critical behavior of a family of $\mathbb{Z}_2$-symmetric scalar field theories on the Bethe lattice (the tree limit of regular hyperbolic tessellations) using both the non-perturbative Functional Renormalization Group and lattice perturbation theory. The family is indexed by the parameter $\z \in (0,1]$, which determines the range of the theory via the kinetic term constructed from the graph Laplacian raised to the power $\zeta$. Specifically, $\z=1$ is the short-range theory, while $0<\z<1$ defines the long-range model. Due to the hyperbolic nature of Bethe lattices, the Laplacian lacks a zero mode and exhibits a spectral gap. We find that upon closing this spectral gap by a modification of the Laplacian, the scalar field theories exhibit novel critical behavior in the form of non-trivial fixed points with critical exponents governed by $\zeta$ and the spectral dimension $d_s=3$. In particular, our analysis indicates the presence of a Wilson-Fisher fixed point for the short range $\zeta =1$ theory. In contrast, the nearest‐neighbor Ising model on the Bethe lattice is known to exhibit mean‐field critical exponents. To the best of our knowledge, this work provides the first evidence that a scalar $\phi^4$ theory and the discrete Ising model on the same underlying lattice may lie in distinct universality classes.

\end{abstract}

\hrule\bigskip
\tableofcontents

%%%%%%%%%%%%%%%%%%%%%%%%%%%%%
\section{Introduction }
\label{sec:intro}
%%%%%%%%%%%%%%%%%%%%%%%%%%%%%
It is well known that on flat Euclidean lattices, the critical behavior of a statistical model is determined by its symmetry and dimensions; for instance, the Ising model and scalar $\phi^4$ theory, both with $\mathbb{Z}_2$ symmetry, share the same critical exponents. In this paper, we explore how curvature can influence critical behavior in the setting of lattice models. A particularly interesting class of lattices are those with uniform negative curvature or regular hyperbolic tessellations. Hyperbolic lattices \cite{Boettcher:2021njg,Maciejko:2020rad} are becoming increasingly relevant in the study of exotic phases of matter, table-top simulations of quantum and statistical systems in curved space-(time) via circuit QED \cite{Boettcher:2019xwl,Kollar:2019ngc,xu2025scalablesuperconductingcircuitframework,Dey:2024jno,Bienias:2021lem}, discrete realization of holographic principle \cite{Yan:2018nco,Asaduzzaman_2020,Okunishi_2024,Erdmenger:2024jsb}, etc. 

\begin{figure}[h]
     \centering
\includegraphics[width=0.35\linewidth]{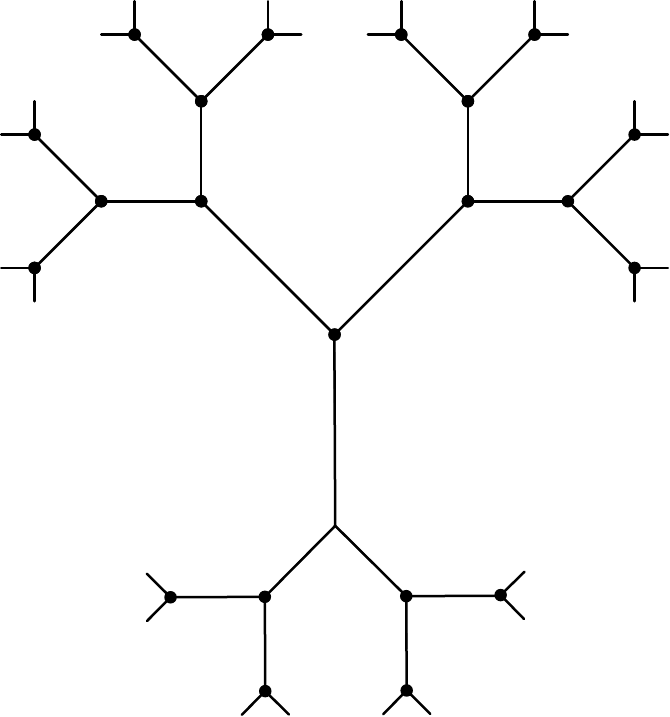}
     \caption{A truncation of a degree-$3$ Bethe lattice $\mathbb{T}_3$ or $\{\infty,3\}$ tesselation of the hyperbolic plane.}
     \label{fig:BL}
 \end{figure}

As a natural starting point for such an investigation, in this paper we focus on the order-$q$ apeirogonal tiling of the hyperbolic (hyper-)plane \cite{PhysRevA.46.1859} or the $q$-regular infinite tree ($q\geq 3$) denoted by $\T_q$, also known as the Bethe lattice. For a representative example see Fig. \eqref{fig:BL}. The hyperbolic nature of these trees has already been exploited to construct discrete toy models of AdS/CFT \cite{Freund:1987kt, Zabrodin:1988ep, Heydeman:2016ldy, Gubser_2017,Gubser_2019,Qu:2021huo, Yan:2023lmj, Okunishi:2023syy,Okunishi_2024}. Moreover, the Bethe lattice offers a playground for exploring the role of different notions of dimension in critical phenomena. Indeed, the concept of `dimension' for a Bethe lattice is far more subtle than in Euclidean lattices. In the latter case, the Hausdorff dimension, spectral dimension, and topological dimension all coincide yielding an unambiguous definition. On the other hand for a $q\geq 3$ degree Bethe lattice, the Hausdorff dimension $d_H = \infty$ \cite{tsuchiya1978spatial}, the spectral dimension $d_s =3$ \cite{Monthus_1996,Delporte:2023saj}, while the the topological dimension is $d_t = 1$ (reflecting the underlying tree structure).

%The concept of dimensionality of the Bethe lattice is more involved than that of Euclidean lattices. Given $q\geq 3$, the number of sites inside a ball of radius $r$ scales exponentially as $(q-1)^r$, indicating that the Hausdorff dimension $d_H$ is infinite \cite{tsuchiya1978spatial}, meanwhile, the scaling of density of states at the lower edge suggests that the {\itshape spectral dimension} $d_s =3$ \cite{Monthus_1996,Delporte:2023saj}. Additionally, the topological dimension is $d_t = 1$, reflecting the underlying tree structure of the Bethe lattice. Apart from serving as a toy model for hyperbolic lattices, the Bethe lattice offers a playground for exploring the role of different notions of dimensions in critical phenomena. 

Thermodynamic quantities and critical exponents for classical spin models are exactly computable on the Bethe lattice \cite{PhysRevLett.74.809,Baxter:1982zz, Hu_1998}. Notably, deep inside the bulk of the Bethe lattice, the Ising model exhibits spontaneous magnetization below a critical temperature $T_c = 2/\log \left( \tfrac{q}{q-2} \right) $, with mean-field critical exponents $\delta = 3$ and $\beta = 1/2$. This aligns with the universality class predicted by the Hausdorff dimension. However, the correlation length does not diverge at the critical temperature, but diverges at absolute zero \cite{Eggarter:1974zz, Breuckmann_2020}, aligning with the topological dimension. Beyond the tree limit, the Ising model on general hyperbolic tessellations exhibits similar behavior \cite{Breuckmann_2020, wu2000ising, Rietman1992TheIM,Iharagi_2010, Okunishi_2024,Krcmar_2008,Ueda_2007}, but with a finite critical temperature at which correlation length diverges. Mathematically, it has been shown that there exists a critical temperature $T_c$, characterizing the onset of spontaneous magnetization, and a lower critical temperature $T^{\prime}_c$ corresponding to the divergence of correlation length. For $T^{\prime}_c<T< T_c$, there exists an intermediate phase where the ordered and disordered phases coexist\cite{wu2000ising}. Numerical studies have suggested that the transition at $T_c$ shows mean-field values for $\beta$ and $\d$, similar to that of the Bethe lattice. The analysis of the intermediate to ferromagnetic phase transition at $T^{\prime}_c$ is quite challenging due to system sizes \cite{Breuckmann_2020} and has been observed numerically very recently \cite{wang2025emergenceboundarysensitivephasehyperbolic}. This differs markedly from flat lattices, where these two temperatures coincide. This difference arises from the exponential growth of volume in hyperbolic lattices, which can counteract an exponentially decaying correlation and cause susceptibility to diverge at $T_c$. In contrast, for a flat lattice a power-law decay of correlation is required to achieve diverging susceptibility.

From the point of view of field theory, there has been an increasing interest in QFT in hyperbolic or Euclidean Anti-de Sitter (EAdS) space \cite{Carmi:2018qzm,Giombi:2025pxx,Copetti:2023sya,Ciccone_2024}. In EAdS, the Laplacian operator has a gapped spectrum, which induces distinct infrared (IR) behavior in comparison to flat spaces \cite{Callan:1989em}. In particular, continuous symmetries can be spontaneously broken in $d=2$ EAdS, exhibiting novel phase transitions in comparison to their flat counterparts. Considerable progress has been made in the analysis of phase transitions in EAdS with diverse methods such as large-$N$ expansion  \cite{Mnasri2015},  conformal bootstrap \cite{Carmi:2018qzm}, Hamiltonian truncation \cite{Hogervorst:2021spa}, functional renormalization group (FRG) \cite{Benedetti:2014gja}, and lattice QFT \cite{Brower:2019kyh, Brower:2022atv}.  Despite the advancements, understanding the phase transitions of field-theoretic models in hyperbolic spaces lacks a general consensus. Evidence for the existence of a critical point for $\phi^4$ theory on hyperbolic lattices has been observed via numerical simulations \cite{Brower:2019kyh, Brower:2022atv}, but computation of the critical exponents has remained elusive. In continuous settings, FRG studies found a mean-field-like transition for scalar $\phi^4$, but analysis in \cite{Mnasri2015} found novel non-mean-field behavior using large-$N$ techniques with distinct $1/N$ corrections from flat space. 

This suggests that further studies are necessary to gain a deeper understanding of critical phenomena in hyperbolic spaces. As a step in this direction, we study the one-component scalar field theory with $\mathbb{Z}_2$ symmetry on the Bethe lattice, via a lattice version of non-perturbative functional renormalization group \cite{Machado:2010wi} in the local potential approximation (LPA) and lattice perturbation theory of long-range models at one-loop. The procedure is implemented using the spectral representation of the Laplacian. 

We observe that the spectrum of the Laplacian is gapped, aligning with the hyperbolic nature of the Bethe lattice. Our key finding is that tuning to the gapless regime yields access to non-Gaussian fixed points. In the absence of such tuning, the gap is a non-running scale in the theory and therefore does not lead to an autonomous flow equation through the rescaling of variables in the FRG framework. In the presence of the gap, the perturbative analysis gives rise to only tree-level beta functions.

 The presence of a non-Gaussian fixed point contrasts with the Ising model on the Bethe lattice, which exhibits a phase transition with mean-field critical exponents. 
 The difference arises because tuning to the gapless theory by addition of a local quadratic term to the Ising model is not possible, unlike the scalar field theory. 
 Any even powers of the Ising spin would evaluate to one, thus, only changing the Hamiltonian by a constant. A closely related line of research studies critical systems on networks with a similar focus on the spectrum of the Laplacian \cite{Erzan,Caldarelli_2024}, but since they do not consider the limit to the gapless regime, the only stable fixed point they find is the Gaussian one. 

We organize the remaining work as follows.
In Section \ref{sec:FTB}, we review the relevant aspects of spectral properties of the graph Laplacian and long-range scalar field theory on the Bethe lattice. In Section \ref{sec:funRG}, we set up the FRG analysis for scalar field theory in the local potential approximation. In Section \ref{sec:pertRG1}, we give further evidence for the existence of a non-Gaussian fixed point by adapting the MS scheme for lattice field theory using a long-range $\epsilon$ expansion scheme, following the approach of \cite{Benedetti:2020rrq, MaPhysRevLett.29.917,Yamazaki:1977pt}. 
In Section \ref{sec:results}, we summarize and discuss our results and propose future research directions.

%%%%%%%%%%%%%%%%%%%%%%%%%%%%%
\section{Field theory on the Bethe lattice}
\label{sec:FTB}
%%%%%%%%%%%%%%%%%%%%%%%%%%%%%

We begin by setting our notation and conventions for the rest of the paper. Our main focus will be a family of scalar field theories  on  the Bethe lattice $\T_q$, indexed by the parameter $\zeta\in (0,1]$,  with bare action (in graph units)
\begin{equation} \label{eq:BLgraphaction}
  S[\chi]= \frac{1}{2}\sum_{i,j}  \chi(i)\left(\Delta_\g^\z\right)_{i,j}\chi(j) + \sum_i\Big\{ \frac{1}{2} m_B^2 \chi(i)^2+\frac{g_B}{4!} \chi(i)^4\Big\}\,.
\end{equation}
With $V(\T_q)$ denoting the vertex set of of the graph,  $i \in V(\T_q)$ labels the vertices, $\chi : V(\mathbb{T}_q) \to \mathbb{R}$ is a scalar field defined on the vertices, and $\D_\g$ is the combinatorial graph Laplacian defined by the following matrix
  \begin{equation} \label{eq:Laplacian}
  \D_\g (i, j)\coloneqq 
\begin{cases}\operatorname{deg}\left(i\right) & \text { if } i=j \\ -1 & \text { if } i \neq j \text { and } i \text { is adjacent to } j \\ 0 & \text { otherwise }
\end{cases}\,.
\end{equation}
The graph Laplacian can be rendered self-adjoint (on a suitable dense domain) in $L^2(\T_q)$, and it has a positive purely continuous spectrum ${\rm spec}(\D_\g)=[\gamma^2,\Lambda_\gamma^2]$, with strictly positive lower bound  $\g^2 = q - 2 \sqrt{q-1}$, and   upper bound $\L^2_{\g} = q + 2 \sqrt{q-1}$.\footnote{ We note that in the short range $\zeta = 1$ case setting, the spectral gap of $\D_\g$ may be regarded as a tree graph analog \cite{Gubser_2017} of the Breitenlohner-Freedman bound \cite{Breitenlohner:1982bm,Breitenlohner:1982jf} in Euclidean Anti-de Sitter space (i.e. hyperbolic space).} The operator $\Delta_\gamma^\zeta$ in \eqref{eq:BLgraphaction} is then defined through the spectral theorem for $\zeta\in (0,1)$. The parameter $\zeta$ determines the range of the theory, with $\z=1$ corresponding to the standard short-range case, while $\zeta\in (0,1) $ yields a long-range model. 
The associated free local density of states (LDOS)  \cite{economou2006green,Delporte:2023saj} 
\begin{equation} \label{eq:LDOSV}
\rho_\g(\ell^2) = \frac{q}{2\pi} \frac{\sqrt{\L^2_\g-\ell^2}\sqrt{\ell^2-\g^2}}{\ell^2(2q-\ell^2)}\,,\quad \ell \in [\gamma, \Lambda_\gamma]\,,
\end{equation}
is normalized such that \footnote{Setting $\lambda= \ell^2 $ recovers the more familiar  normalization 
$
\int_{\g^2}^{\L_\g^2}	 \rho_\g(\l) d\l
 = 1\,.$ 
Treating $\rho$ as a function of  $\ell^2$ is notationally convenient for the Functional Renormalization Group analysis in Section \ref{sec:funRG} .
}
\begin{equation} \label{eq:Dosnorm}
 2 \int^{\L_\g}_{\g} \ell \, \rho_\g(\ell^2) d\ell = 1\,. 
\end{equation}
 \begin{figure}[h]
     \centering
\includegraphics[width=0.5\linewidth]{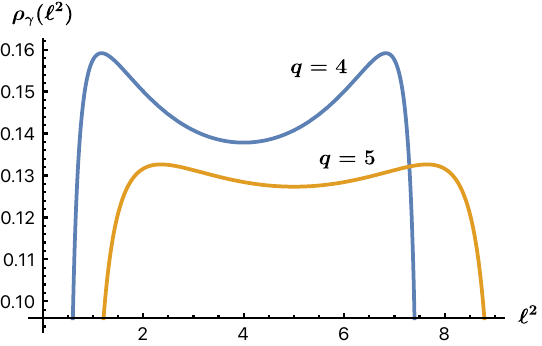}
     \caption{Local density of states of the $\T_4$ (blue) and $\T_5$ (orange) Bethe lattices.}
     \label{fig:LDOS}
 \end{figure}
Close to the lower bound of the spectrum, $\rho_\g(\ell^2)$ has the leading  scaling behavior
 \begin{equation} \label{eq:LDOSscaling}
  \rho_\g(\ell^2) \propto  (\ell^2-\g^2)^{\frac{d_s-1}{2}}\,,
\end{equation}
where the exponent $d_s$ is referred to as the {\itshape spectral dimension}; it follows readily from \eqref{eq:LDOSV} that $d_s = 3$ for any $\T_q$.
The spectral gap $\g^2$ acts a natural IR regulator and may be regarded as an additional scale in the scalar field theory.

Our main result, detailed in Sections \ref{sec:funRG} and \ref{sec:pertRG1}, is the appearance of non-trivial critical behavior when the spectral gap is tuned to zero through a redefinition of the Laplacian by a constant shift. To that end, it is  convenient to define a `gapless' Laplacian $\D_0 \coloneqq  \D_\g - \g^2 \mathbf{1}$ with ${\rm spec}(\Delta_0) = [0,\L_0^2]$, where $\L_0^2 \coloneqq \L^2_\g - \g^2$. The corresponding LDOS $\rho_0(\ell^2)$ can be simply obtained by substituting $\ell^2 \to \ell^2 + \g^2$ in $\rho_\g(\ell^2)$ and is given by
 \begin{equation} \label{eq:Gaplless}
 \rho_0(\ell^2) = \frac{q}{2\pi} \frac{\sqrt{\ell^2} \sqrt{\L_0^2-\ell^2} }{(\ell^2+\g^2)(\L^2_\g-\ell^2)}\,, \quad \ell\in [0,\Lambda_0]\,.
\end{equation}
Further, for brevity of notation, we will use $\Delta$ to denote both $\Delta_\gamma$ and $\Delta_0$, in Sections \ref{sec:funRG} and \ref{sec:pertRG1} whenever the discussion applies to both cases. We introduce $\sigma(\ell)$ to denote the domain of the spectral integrals, which is the closed interval $[\gamma,\Lambda_\gamma]$ for $\Delta_\gamma$ and $[0,\Lambda_0]$ for $\Delta_0$.

%%%%%%%%%%%%%%%%%%%%%%%%%%%%%
\section{Functional Renormalization Group on the Bethe lattice}
\label{sec:funRG}
%%%%%%%%%%%%%%%%%%%%%%%%%%%%%

The Functional Renormalization Group (FRG) is one of the most widely and fruitfully used techniques in quantum many body physics, favored for its ability to efficiently identify fixed points and compute critical exponents. We present a schematic overview of the FRG, tailored to our setting, referring the reader to  the articles \cite{Banerjee:2018pkt,Machado:2010wi,Dupuis_2008,Caillol:2012dt,Caillol:2013eta} and monographs \cite{Wipf:2013vp,kopietz2010introduction} for further details. 

The FRG is defined by introducing a scale $k$ dependent mode modulation term (with ``$\cdot$'' denoting a sum over vertices)
\begin{equation} \label{eq:regulator}
  \D S_{k}[\chi] = \frac{1}{2}  \chi \cdot R_k(\D) \cdot \chi\,,
\end{equation}
to the bare action, yielding the mode-modulated action  $S_k \coloneqq S + \D S_k$. The regulator kernel $R_k$ is chosen to suppress the low-energy eigenmodes of the Laplacian below scale $k$, and vanish at $k=k_{\min}$, where $k_{\rm min}^2$ is the bottom of the Laplacian's spectrum. The scale $k$ then interpolates between the bare and renormalized theories.

Performing the functional integral yields the mode-modulated free energy $W_k$, 
\begin{equation} \label{eq:Zk}
 e^{W_k[J]} = \int \DD \chi \exp{- S_k[\chi] +  J \cdot \chi},
\end{equation}
whose (modified) Legendre transform 
\begin{equation} \label{eq:EAAdef}
  \G_k[\phi] \coloneqq - W_k [J[\phi]] +  J[\phi] \cdot \phi - \D S_k[\phi]\,,\qquad \frac{\d W[J]}{\d J}\bigg|_{J=J[\phi]}\overset{!}{=}\phi\,,
\end{equation}
satisfies the functional integro-differential equation
\begin{equation} \label{eq:FRGE}
  k\p_k \G_k[\phi] = \frac{1}{2} \Tr{k \p_k R_k \left[ \G_{k}^{(2)}[\phi] + R_k \right]^{-1} }\,, \quad \G_{k}^{(2)}[\phi]:= \frac{\delta^2 \Gamma_k}{\delta \phi \delta \phi}\,,
\end{equation}
known as the Wetterich  equation or the Functional Renormalization Group equation (FRGE). 

Next, we work with the so-called optimized regulator
\begin{equation} \label{eq:litimcutoff}
  R_k(\D) \coloneqq (k^{2\zeta}-\D^{\z})\Theta(k^{2\zeta}-\D^{\z})\,,
\end{equation}
which has the desirable feature that for scales $k\geq k_{\rm max}$ (where  $k_{\rm max}^2$ is the upper bound of the Laplacian's spectrum) all propagating modes are frozen out, and 
the functional integral in \eqref{eq:Zk} becomes a product of single site integrals. It then follows from \eqref{eq:EAAdef} that $\Gamma_k$ has the form
\begin{equation} \label{eq:ulexact}
  \G_k[\phi] = \frac{1}{2}  \phi\cdot\D^\zeta\cdot \phi+
  \sum_i U_{k}(\phi(i))\,,\qquad k^2\geq \Lambda^2\,,
\end{equation}
with the potential $U_k$ satisfying an integro-differential equation \eqref{eq:singsite}.

For $k<k_{\rm max}$, solving \eqref{eq:FRGE} requires some truncation ansatz for $\Gamma_k$. In this paper, we work with the Local Potential Approximation (LPA) ansatz
\begin{equation} \label{eq:LPAansatz}
  \G_k[\phi] = \frac{1}{2}  \phi\cdot\D^\zeta\cdot \phi+
  \sum_i U_{k}(\phi(i))\,,
\end{equation}
where the dependence on the scale $k$ is completely contained in the effective potential $U_k(\phi)$. For systems with small anomalous dimensions, the LPA yields fairly accurate results, with the critical values of the coupling being consistent with those obtained from other methods \cite{Caillol:2012dt,Caillol:2013eta,Banerjee:2018pkt}.  

Inserting \eqref{eq:LPAansatz} into the FRGE \eqref{eq:FRGE} and specializing to homogeneous field configurations $\phi\equiv  \phi_0$ leads to the flow equation for the effective potential 
\begin{equation} \label{eq:LPAflow} 
  k \p_k U_k(\phi_0) = \int_{\s(\ell) }
  d\ell\,\ell\rho(\ell^2) 
  \frac{k\p_k R_k(\ell^2)}{\ell^{2\z} + R_k(\ell^2)+ U_k^{(2)}(\phi_0)}\,,
\end{equation}
with the integral with respect to the LDOS $\rho(\ell^2)$ arising from the functional trace (upon omitting a volume term), and $U_k^{(2)} \coloneqq \partial ^2 U_k/\partial\phi_0^2$.
Finally, due to the step function in the regulator \eqref{eq:litimcutoff}, the flow equation reduces to
\begin{equation}\label{eq:optimizedLPAflow}
	k \p_k U_k(\phi_0) = \frac{ \z k^{2\z}}{k^{2\z}+ U^{(2)}_k(\phi_0)}\text{Vol}(k)\, , 
	\quad \text{Vol}(k)\coloneqq 2\int_{\s(\ell)} d\ell\,\ell \rho(\ell^2)
	\Theta(k^{2\z}-\ell^{2\z})\,.
\end{equation}

%%%%%%%%%%%%%%%%%%%%%%%%%%%%%
\subsection{Critical exponents}
\label{sec:cbasis}
%%%%%%%%%%%%%%%%%%%%%%%%%%%%%

The above flow equation \eqref{eq:optimizedLPAflow} cannot be made autonomous as $\text{Vol}(k)$ does not have a scaling form for general $k$. Nevertheless, one might find a set of variable transformations in the low energy scaling limit of the LPA flow equation \eqref{eq:optimizedLPAflow} i.e., $k \to k_{\min}$. In this regime, using \eqref{eq:LDOSscaling}, we find that 
\begin{equation} \label{eq:Vscalebare}
\text{Vol}(k) \sim \frac{c_q}{\z} (k^2-k^2_{\min})^{d_s/2} \,, \quad c_q \coloneqq \frac{2 \z q(q-1)^{1/4} }{3\pi  (q-2)^2}\,.
\end{equation}
Here the symbol $g(x) \sim f(x)$ as $x \to x_0$, denotes that $\lim_{x\to x_0}\frac{f(x)}{g(x)} = 1$. In the gapped case, $k_{\min} = \gamma$, the gap $\gamma$ is a non-running intrinsic scale, and a suitable rescaling of variables leading to an autonomous equation cannot be found. This is a generic feature of FRG with a non-running scale, such as at finite temperature \cite{Berges:2000ew}, in non-commutative space-time \cite{Gurau:2009ni}, and has been observed in FRG for the continuum hyperbolic space \cite{Benedetti:2014gja}.

However, for the gapless Laplacian $\D_0$, an appropriate change of variables does exist that leads to an autonomous flow equation. 
In this case, $k_{\min} = 0$ and $k_{\max} = \L_0$. As $k \to 0$ we find
\begin{equation} \label{eq:Vscale}
\text{Vol}(k) \approx \frac{c_q}{\z} k^{d_s}\,.
\end{equation}

 To transition to an autonomous LPA flow we define the following transformations
\begin{equation} \label{eq:auttransform}
  V_k(\vph) \coloneqq \frac{1}{c_q k^{d_s}}U_k(\phi_0(\vph))\,,\quad 
 \vph(\phi_0)\coloneqq \frac{k^{\frac{2\z-d_s}{2}}}{\sqrt{c_q}}\,\phi_0\,,\quad
 v(\t) \coloneqq \frac{\text{Vol}(\L_0 \t)}{\frac{c_q}{\z} (\L_0 \t)^{d_s}}\,.
\end{equation}
The function $v(\t)$ is a conveniently defined normalized volume such  that $v(0)=1$ and $v(1) = \frac{1}{c_q \L_0^{d_s}}$, where we have defined a normalized scale $\t  \coloneqq \frac{k}{\L_0}$. We obtain a closed form expression for $v(\t )$ by computing the integral in \eqref{eq:optimizedLPAflow} and using definition \eqref{eq:auttransform}
\begin{equation} \label{eq:normvol}
  v(\t ) = \frac{\pi-q \arcsin\left(1-2 \t^2\right)-(q-2) \arctan\left(\frac{(q-2) \left(2 \t^2-1\right)}{2 q \t \sqrt{1-\t^2}}\right) }{2 \pi c_q (\L_0 \t)^{d_s} }\,.
\end{equation} 
In the above-defined variables, we get the flow equation
\begin{equation} \label{eq:autFP}
  \t  \partial_\t  V_\t (\vph)  =\frac{d_{s}-2\zeta}{2} \varphi V_{\t }^{(1)}(\vph) -d_{s} V_\t (\vph)+\frac{v(\t )}{1+  V_{\t }^{(2)}\,(\vph)}\,,
\end{equation}
which becomes autonomous for small $\t$, independent of value of $q$. In the regime $\t \to 0$, \eqref{eq:autFP} coincides with the continuum flow equation for the one-component scalar field theory in $d_s$ dimensions,
\begin{equation} \label{eq:autFP2}
  \t \partial_\t  V_\t(\vph)  =\frac{d_{s}-2\zeta}{2} \varphi V_{\t}^{(1)}(\vph) -d_{s} V_{\t}(\vph)+\frac{1}{1+  V_{\t}^{(2)}\,(\vph)}\,,
\end{equation}
with the associated fixed point equation
\begin{equation} \label{eq:FPeq}
  \frac{d_{s}-2\zeta}{2} \varphi V^{*}{}^{(1)}(\vph) -d_{s} V^{*}(\vph)+\frac{1}{1+  V^{*}{}^{(2)}\,(\vph)} =0\,,
\end{equation}
where $V^{*}(\vph)$ is the fixed point potential.
Inserting the Taylor series $V_{\t }(\vph) = \sum_{i \geq 0} \frac{g_{2 i}(\t)}{(2 i)!} \vph^{2 i}$ in \eqref{eq:autFP} leads to the beta functions $\b_{2i}$ of the couplings $g_{2i}$
\begin{equation} \label{eq:autbeta}
  \t  \partial_\t  g_{2 i}=\beta_{2 i}\left(g_2, \ldots, g_{2 i+2}\right), \quad i \geq 1\, ,
\end{equation}
which leads to an infinite system of coupled ODEs. To make the computations tractable, we implement a truncation scheme such that $g_{2i+2} = 0$ for all $i>N$, where $N$ is the order of truncation.
Thus, \eqref{eq:autbeta} reduces to a closed system of $N+1$ ODEs,
\begin{equation} \label{eq:tansatz}
   \t  \partial_\t  g_{2 i}=\beta_{2 i}\left(g_2, \ldots, g_{2 i+2}\right)\,, \quad 1 \leq i \leq N\,.
\end{equation}

 We find the fixed point solutions $g^{*}_{2i}$ by setting L.H.S of \eqref{eq:tansatz} to zero (in the small $\t$ regime), and solving the resulting  algebraic equation. The associated critical exponents $\th_{j,\z}$ (subscripts $j,\z$ denote the $j$th eigenvalue for given $\z$) can be determined by computing the eigenvalues of the stability matrix
\begin{equation} \label{eq:stabmat}
  M(g^{*})_{ij} \coloneqq\left.\frac{\partial \beta_{2 i}}{\partial g_{2 j}}\right|_{g=g^*}\,.
\end{equation}
In general, the truncated beta functions \eqref{eq:tansatz} produces many spurious fixed points. However, apart from the Gaussian fixed point, we find the `true' solution by requiring that the corresponding stability matrix possesses only one negative eigenvalue.
 We performed the computation up to a truncation of $\mathcal{O}(16)$. We direct our attention to two particular cases of $\z$. One, the short-range theory, i.e., $\z=1$ and the Gaussian limit $\z= 3/4$, about which we perform our perturbative analysis in Section \ref{sec:pertRG1}. For the short-range theory, we find the Gaussian and the Wilson-Fisher fixed points. The negative real critical exponent corresponding to the relevant direction of the Wilson-Fisher fixed point is $\th_{1,1} \approx -1.541$. For $\z = \frac{3}{4}$, the only `true' fixed point is Gaussian and the critical exponents are then given by canonical dimensions of the couplings in continuum $3$D, with one negative critical exponent $\th_{1,\,3/4} = -\frac{3}{2}$, corresponding to the quadratic coupling. 
To check for consistency with perturbation theory (discussed in Section \ref{sec:pertRG1}), we obtain the critical exponents for $\z = \frac{3+\epsilon}{4}$ for small $\epsilon$ at $\mathcal{O}(4)$ truncation, analytically. They are given by $\{\th_{1,\,(3+\epsilon)/4},\th_{2,\,(3+\epsilon)/4}\} = {\left\{-\frac{3}{2}-\frac{\epsilon }{6},\epsilon \right\}}$.

%%%%%%%%%%%%%%%%%%%%%%%%%%%%%
\subsection{Critical line of the gapless theory}
\label{sec:critline}
%%%%%%%%%%%%%%%%%%%%%%%%%%%%%
The critical line of the theory is the set of value of bare parameters such that the theory flows to the fixed point. We obtain it for the gapless theory via two complementary methods in the framework of the FRG. First, we solve the truncated beta functions obtained in \eqref{eq:autbeta}. The initial data of the system of ODEs can be computed exactly using \eqref{eq:singsite} and \eqref{eq:auttransform}. Second, we integrate directly \eqref{eq:optimizedLPAflow} without using any polynomial truncation, but in this approach, we approximate the initial data with the bare potential at very large $k$. Both methods and a comparison of their results for $\z=1$ and $\z= 3/4$ are described below.

%%%%%%%%%%%%%%%%%%%%%%%%%%%%%
\paragraph{\underline{Critical line in coupling basis}}\mbox{}\\
%%%%%%%%%%%%%%%%%%%%%%%%%%%%%

For small perturbations $\d g(\t)$ around the fixed point $g^{*}$, the flow trajectory would be dominated by the relevant direction described by the unique linear combination 
\begin{equation} \label{eq:LinUnstabManifold}
  \sum^{N}_{i=1} a_i \d g_{2i} = c\,\t^{\th_{1,\z}}\,, \quad 0<\t\ll1\,.
\end{equation}
The constant $c =0$ describes the linearized unstable manifold, i.e., the co-dimension one hyperplane along which couplings flow to the fixed point. The critical line is evaluated by determining the set of initial conditions that flow to the linearized unstable manifold.

%The critical line is parametrized by $\{m^2_B,g_{B}\}$ such that with this choice, the couplings flow to the linearized unstable manifold. 

 In the coupling basis, the initial conditions are computed at the ultra-local scale $\L_0$, where the effective average action at that scale $\G_{\L_0}$ can be calculated from single-site integrals. To obtain $\G_{\L_0}$, we define the scale-dependent effective action $\Gh_k[\phi] \coloneqq \G_k[\phi] + \D S_k[\phi] $ and we note from the definition of effective action that, 
\begin{equation} \label{eq:integrodiffeq}
  e^{-\Gh_k[\phi]+ \Gh_{k}^{(1)}[\phi]\cdot \phi} = \int \DD \chi \exp{- S[\chi]+  \frac{1}{2}  \chi \cdot R_k(\D_0) \cdot \chi +  \Gh_{k}^{(1)}[\phi] \cdot \chi}\,.  
\end{equation}
For $k^{2\z}  =\L_0^{2\z}$, \eqref{eq:integrodiffeq} factorizes into single site integrals \begin{equation} \label{eq:singsite}
e^{-\hat{U}_{\L_0}(\phi_0)+ \phi_0 \,\hat{U}_{\L_0}^{(1)}(\phi_{0})}=\int^{\infty}_{-\infty} d \chi \, e^{-\frac{1}{2}\left(m_B^2+\L_0^{2\zeta }\right) \chi^2-\frac{g_{B}}{4!} \chi^4+\chi\,\hat{U}_{\L_0}^{(1)}(\phi_{0}) }\,.
\end{equation}
Following the definition $\hat{\G}_k$, at the ultra-local scale, we get the corresponding scale-dependent effective potential to be $\hat{U}_{\L_0}(\phi_0) \coloneqq U_{\L_0}(\phi_0) + \frac{1}{2} \L^{2\z}_0 \phi_0^2$.

We use the transformation rules \eqref{eq:auttransform} and substitute the truncated coupling basis \eqref{eq:tansatz} into \eqref{eq:singsite} to obtain the initial conditions $\{g_{2i}(\t=1)\}$ in terms of bare parameters $m^2_B$ and $g_B$.  In Appendix \ref{app:ULC}, we show that the large-$q$ limit corresponds to the Gaussian theory in the LPA.

With initial data in hand, we integrate the beta functions in \eqref{eq:autbeta}. We obtain the critical line for the short-range theory in the hopping parameterization \cite{montvay1994quantum} of the bare action. In this form the action is given by
\begin{equation} \label{eq:SHop}
  S[\tilde{\chi}] = -\k\sum_{(i,j)} \tilde{\chi}(i) \tilde{\chi}(j)+\sum_i \left\{\tilde{\chi}^2(i) + \l \left(\tilde{\chi}^2(i)-1\right)^2 \right\}\,,
\end{equation}
% {\color{blue}{
where $\kappa$ is the hopping parameter and $\lambda$ is the coupling constant,
% }}
and the new parameters of the action are related to the bare quadratic and quartic couplings by
% {\color{blue}{
% to the old ones 
% }}
% by
\begin{equation} \label{eq:hoppingaction}
  m_B^2 = \frac{1-2 \lambda }{\kappa }-q+2 \sqrt{q-1}\,,\quad g_B = \frac{6 \lambda }{\kappa ^2}\,, \quad \chi = \sqrt{2\k} \tilde{\chi}\,.
\end{equation}
To find the critical value of $\k_c(\l)$ numerically, we fix $\l$, and tune to a value of $\k$ such that the flow trajectories reach close to the linearized unstable mainfold at $\t = 0.001$. For the long-range theories, the hopping parametrization does not exist, and the critical line is obtained in terms of the bare couplings. Accordingly, we fix $g_B$ and determine the critical quadratic coupling 
$m^2_{B,c}(g_B)$ by requiring the flow to approach the linearized unstable manifold at $\tau = 0.001$. 

\begin{figure}[H]
    \centering
    \includegraphics[width=0.625\linewidth]{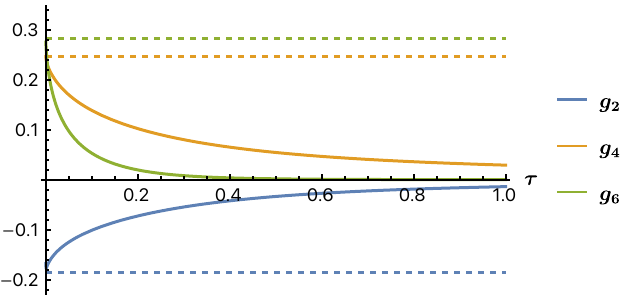}
    \caption{Critical flow of couplings $g_2,\,g_4$ and $g_6$ for $\T_{5}$ at $\mathcal{O}(16)$ truncation for short range, $\z=1$. For the short-range model, the couplings flow towards the interacting fixed point from critical initial conditions. The fixed point value of the couplings is denoted by dashed lines of the corresponding color in the figure.} 
    \label{fig:coupflow_1}
\end{figure}

\begin{figure}[H]
    \centering
    \includegraphics[width=0.625\linewidth]{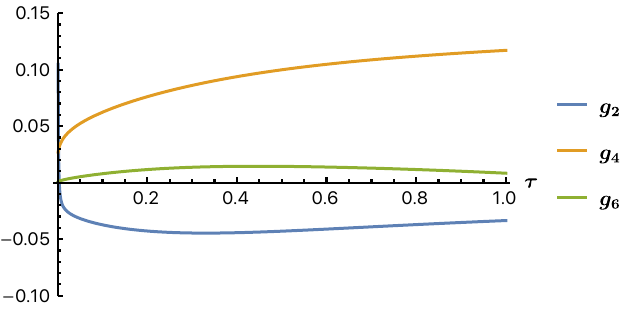}
    \caption{Critical flow of couplings $g_2,\,g_4$ and $g_6$ for $\T_{5}$ at $\mathcal{O}(16)$ truncation for long-range at $\z= 3/4$. The couplings flow towards the Gaussian fixed point.}
    \label{fig:coupflow_2}
\end{figure}

In this work, we implemented the computation of the critical line in $\mathcal{O}(16)$ truncation for $q = 3,5,6$ and $7$ and up to $\l = 1.0$ for $\z=1$, and $g_B=15.0$ for $\z= 3/4$. The numerical values of the critical couplings are tabulated in Table \ref{tab:critline} and \ref{tab:critlinegauss} for $\z=1$ and $\z= 3/4$, respectively. In Figures \ref{fig:coupflow_1} and \ref{fig:coupflow_2}, we have shown a sample critical flow of the couplings $g_2$, $g_4$, and $g_6$ for $q=5$ towards their respective fixed points, for both short- and long-range theories determined by solving the beta functions at $\OO(16)$.

%%%%%%%%%%%%%%%%%%%%%%%%%%%%%
\paragraph{\underline{Critical line without polynomial truncation}}\mbox{}\\
%%%%%%%%%%%%%%%%%%%%%%%%%%%%%

\begin{figure}[H]
    \centering
    \includegraphics[clip, trim= 5cm 10.2cm 5cm 3.6cm, width=0.8\linewidth]{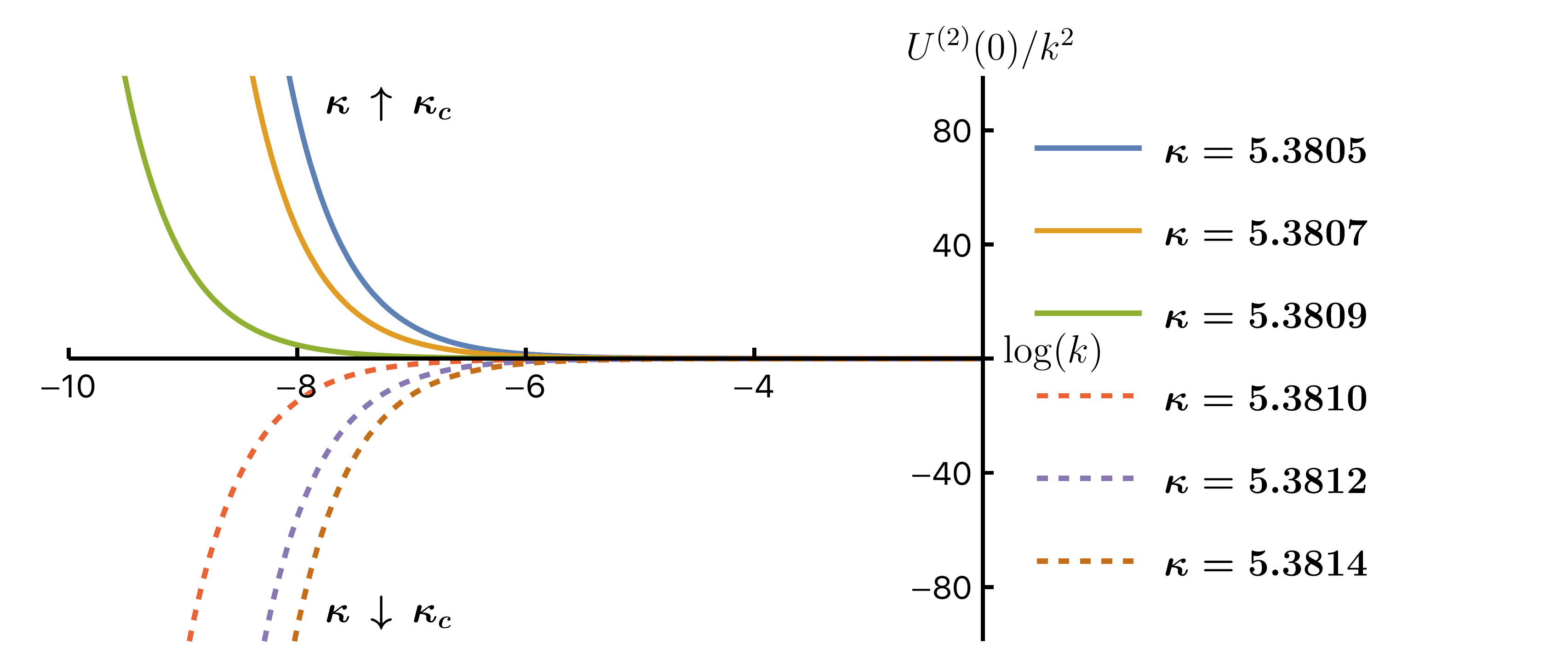}
    \caption{The scaled Hessian $U^{(2)}(0)/k^2$ for $\T_3$ with interaction $\lambda= 0.06$ in the hopping parameterization. The dashed lines correspond to $\k > \k_c$ and solid lines correspond to $\k < \k_c$. Starting from hopping parameter $\k = 5.3805$ denoted by the solid blue line, we increase $\k$ up to $\k = 5.3814$, depicted by dashed brown line. As $k\to 0$, $U^{(2)}(0)/k^2$ changes sign between $\k = 5.3809$ to $\k = 5.3810$, indicating $5.3809 < \k_c(\l=0.06) < 5.3810$.  }
    \label{fig:CriticalPot}
\end{figure}

Alternatively, we also obtain the critical line by 
numerically solving \eqref{eq:optimizedLPAflow}, without implementing a polynomial truncation of the theory. Using the LPA flow equation \eqref{eq:optimizedLPAflow}, we analyze the behavior of the scaled Hessian $U_{k}^{(2)}(0)/k^{2\z}$ subject to different initial conditions. In the unbroken phase, the point $\phi_0 = 0$ is the local minimum of the potential, therefore $U_{k}^{(2)}(0) >0 $. In the broken phase, where $\phi_0$ is a local maximum, then $U_{k}^{(2)}(0) <0$. To determine the critical line, we fix $\l$ and vary $\k$. As $k\to 0$, for $\k>\k_c(\l)$ we are in the broken phase, hence $U_{k}^{(2)}(0) <0$, and for $\k<\k_c(\l)$, we are in the unbroken phase, therefore $U_{k}^{(2)}(0) >0$ as shown in Fig. \ref{fig:CriticalPot}. At a critical point $\k_c(\l)$, the potential becomes flat and $U^{(2)}_k(0) = 0$ as $k \to 0$. Practically, we monitor the sign of $U^{(2)}(0)$ and estimate the value $\k_c(\l)$ to be in between two values of $\k$ where the second derivative $U^{(2)}(0)$ flips sign as described in Fig. \ref{fig:CriticalPot}. 
A similar procedure is utilized to obtain the critical line of the long-range theory in terms of the bare couplings by fixing $g_B$ and varying $m_B^2(g_B)$ to find a critical value of the quadratic coupling $m^2_{B,c}(g_B)$.

However, in the absence of a polynomial truncation ansatz, the ultra-local effective action in \eqref{eq:singsite} is hard to compute. It is therefore useful to start from the $k\to\infty$ limit of the ultra-local action as the initial data.  
To obtain the initial conditions, we look at the integro-differential equation for the effective average action as $k\geq \L_0$ 
\begin{equation} \label{eq:integro}
  e^{-U_k(\phi_0)} = \int d \chi \exp{-U_{B}(\chi) - \frac{1}{2} k^2 (\phi_0-\chi)^2 - (\phi_0-\chi)U_k^{(1)}(\phi_0)}\,,
  \end{equation}
where $U_B$ is the bare potential. For large $k$,  at the leading order, we obtain the mean-field initial data
\begin{equation} \label{eq:MFT}
  U_{k}(\phi_0) \sim U_{B}(\phi_0) =\frac{1}{2} m_B^2 \phi_0^2 + \frac{g_B}{4!} \phi_0^4    \,.
\end{equation}

In practice, we set the initial data to be the mean field at very large $k = k_{\text{max}}\sim e^9$. For a $\Z_2$ symmetric scalar field theory, a similar setup has been used to obtain the critical line on three and four-dimensional (hyper-)cubic lattices with excellent agreement with Monte-Carlo data in \cite{Caillol:2012dt,Caillol:2013eta}. 
Thus for practical implementation the initial condition for 
\eqref{eq:optimizedLPAflow} is then given by
\begin{equation} \label{eq:initcondpot}
  U_{k_{\text{max}}}(\phi_0) = U_{B}(\phi_0)    \,.
\end{equation}
To solve \eqref{eq:optimizedLPAflow} numerically, we constraint $\phi_0$ to lie in a closed interval $\phi_0 \in [-\phi_{\max},\phi_{\max}]$, and implement boundary conditions
\begin{equation} \label{eq:boundcondpot}
  U_{k}(\pm\phi_{\text{max}}) = U_{B}(\pm\phi_{\max}) \,,
  \quad \forall k  \,,
\end{equation}
where we chose $\phi_{\max} = 8$.  

%%%%%%%%%%%%%%%%%%%%%%%%%%%%%
\paragraph{\underline{Results for the critical line}}\mbox{}\
%\label{sec:potflow}
%%%%%%%%%%%%%%%%%%%%%%%%%%%%%

\begin{figure}[H]  
    \centering
    \begin{subfigure}[t]
    {0.48\textwidth}
        \centering
        \includegraphics[width=\linewidth]{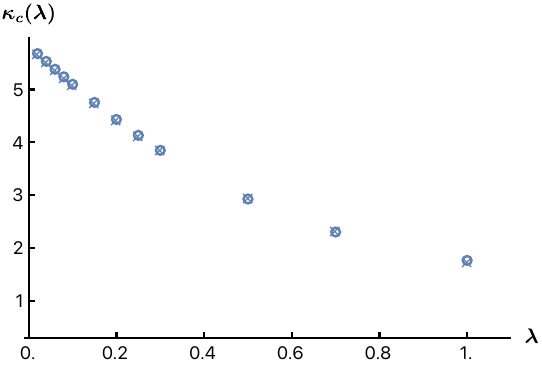} % 
    \end{subfigure}
    \;
    \begin{subfigure}[t]
    {0.48\textwidth}
        \centering
        \includegraphics[width=\linewidth]{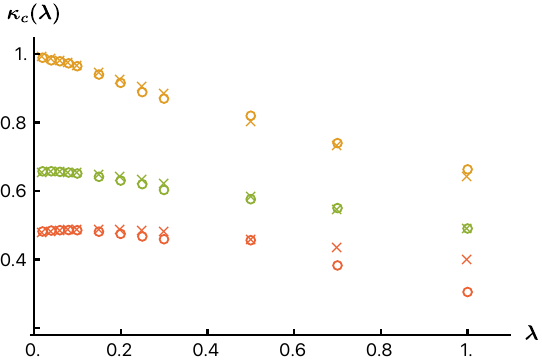} 
    \end{subfigure}
    \caption{The critical line for $\T_3$ in blue (left) and the critical line for $\T_5$, $\T_6$ and $\T_7$ in orange, green and red (right) in the hopping parameterization. The circles correspond to critical points obtained via the truncation method, and the crosses correspond to critical points in the full potential formalism. The critical line for $\T_3$ is separted from the rest for clearer presentation as values of $\k_c(\l)$ for small $\l$ are considerably larger for $\T_3$, when compared to the other cases.}
     \label{fig:CriticalLine12}

\end{figure}

We present the numerical values of $\k_c(\l)$ for the short-range theory via the two methods discussed above in Table \ref{tab:critline} and the corresponding critical line in Figure \ref{fig:CriticalLine12}.
\begin{table}[h!]
\centering
\begin{tabular}{|l|c|c|c||c|c|c||c|c|c||c|c|c|}
\hline
$\lambda$ & $q$ & $\kappa_c^{\text{tr}}$ & $\kappa_c^{\text{di}}$ & 
$q$ & $\kappa_c^{\text{tr}}$ & $\kappa_c^{\text{di}}$ &
$q$ & $\kappa_c^{\text{tr}}$ & $\kappa_c^{\text{di}}$ &
$q$ & $\kappa_c^{\text{tr}}$ & $\kappa_c^{\text{di}}$ \\
\hline
$0.02$ & $3$ & $5.680$ & $5.678$ & $5$ & $0.997$ & $0.996$ & $6$ & $0.659$ & $0.657$ & $7$ & $0.483$ & $0.482$ \\
$0.04$ &     & $5.532$ & $5.530$ &     & $0.992$ & $0.990$ &     & $0.660$ & $0.658$ &     & $0.487$ & $0.484$ \\
$0.06$ &     & $5.381$ & $5.387$ &     & $0.986$ & $0.982$ &     & $0.661$ & $0.656$ &     & $0.489$ & $0.485$ \\
$0.08$ &     & $5.243$ & $5.239$ &     & $0.979$ & $0.973$ &     & $0.660$ & $0.654$ &     & $0.492$ & $0.486$ \\
$0.10$ &     & $5.103$ & $5.097$ &     & $0.972$ & $0.965$ &     & $0.659$ & $0.651$ &     & $0.492$ & $0.485$ \\
$0.15$ &     & $4.764$ & $4.755$ &     & $0.952$ & $0.941$ &     & $0.653$ & $0.642$ &     & $0.493$ & $0.481$ \\
$0.20$ &     & $4.444$ & $4.432$ &     & $0.931$ & $0.916$ &     & $0.646$ & $0.631$ &     & $0.491$ & $0.475$ \\
$0.25$ &     & $4.146$ & $4.146$ &     & $0.910$ & $0.889$ &     & $0.637$ & $0.620$ &     & $0.488$ & $0.467$ \\
$0.30$ &     & $3.868$ & $3.847$ &     & $0.889$ & $0.870$ &     & $0.628$ & $0.603$ &     & $0.487$ & $0.459$ \\
$0.50$ &     & $3.868$ & $3.847$ &     & $0.808$ & $0.820$ &     & $0.588$ & $0.576$ &     & $0.463$ & $0.456$ \\
$0.70$ &     & $2.962$ & $2.926$ &     & $0.736$ & $0.740$ &     & $0.549$ & $0.550$ &     & $0.439$ & $0.382$ \\
$1.00$ &     & $1.762$ & $1.746$ &     & $0.646$ & $0.663$ &     & $0.495$ & $0.490$ &     & $0.404$ & $0.305$ \\
\hline
\end{tabular}
\caption{Numerical values of critical hopping parameter $\kappa_c$ computed in the coupling basis~($\kappa_c^{\text{tr}}$) and from directly integrating the LPA potential ($\kappa_{c}^{\text{di}}$) for $\z= 1$.}
 \label{tab:critline}
\end{table}

From Table \ref{tab:critline} and Fig. \ref{fig:CriticalLine12}, for smaller values $\l$, we observe excellent agreement between both methods for all the values of $q$ considered. The agreement worsens with both increasing $\l$ and $q$, independently. In the beta-function method, increasing $\l$, the numerical values of the couplings, $g_{2i}(1)$, at ultra-local conditions differ by large orders of magnitude, making numerical analysis challenging. As demonstrated in Appendix \ref{app:ULC}, the ultra-local conditions are pushed towards the Gaussian fixed point with increasing $q$, demanding higher precision for solving the differential equations. In the direct integration of the LPA flow equation, there are two distinct factors affecting the accuracy. First, the mean-field initial condition is only approximated by starting the flow at a large but finite ultraviolet scale $k$. Second, for numerical tractability, the field variable is constrained to lie in a finite interval. An interesting future work would be to conduct a more detailed numerical investigation for a larger  range of values of $\l$ and $q$. 

We also present the numerical values of the critical quadratic coupling $m^{2}_{B,c}(g_B)$ for $\z= 3/4$ in Table \ref{tab:critlinegauss}.

\begin{table}[h!]
\centering
\begin{tabular}{|l|c|c|c||c|c|c||c|c|c||c|c|c|}
\hline
$g_B$ & $q$ & $-m^{2\text{ tr}}_{B,c}$ & $-m^{2\text{ di}}_{B,c}$ & 
$q$ & $-m^{2\text{ tr}}_{B,c}$ & $-m^{2\text{ di}}_{B,c}$ &
$q$ & $-m^{2\text{ tr}}_{B,c}$ & $-m^{2\text{ di}}_{B,c}$ &
$q$ & $-m^{2\text{ tr}}_{B,c}$ & $-m^{2\text{ di}}_{B,c}$ \\
\hline
$0.4$ & $3$ & $0.168$ & $0.165$ & $5$ & $0.113$ & $0.112$ & $6$ & $0.102$ & $0.101$ & $7$ & $0.934$ & $0.093$ \\
$0.6$ &     & $0.242$ & $0.234$ &     & $0.166$ & $0.163$ &     & $0.150$ & $0.147$ &     & $0.138$ & $0.136$ \\
$0.8$ &     & $0.313$ & $0.307$ &     & $0.217$ & $0.212$ &     & $0.196$ & $0.192$ &     & $0.181$ & $0.179$ \\
$1.0$ &     & $0.381$ & $0.372$ &     & $0.267$ & $0.260$ &     & $0.242$ & $0.236$ &     & $0.224$ & $0.218$ \\
$1.5$ &     & $0.542$ & $0.527$ &     & $0.386$ & $0.374$ &     & $0.352$ & $0.341$ &     & $0.327$ & $0.316$ \\
$2.0$ &     & $0.693$ & $0.668$ &     & $0.500$ & $0.482$ &     & $0.456$ & $0.440$ &     & $0.425$ & $0.409$ \\
$3.0$ &     & $0.972$ & $0.933$ &     & $0.714$ & $0.688$ &     & $0.656$ & $0.629$ &     & $0.614$ & $0.587$ \\
$4.0$ &     & $1.228$ & $1.188$ &     & $0.917$ & $0.877$ &     & $0.844$ & $0.801$ &     & $0.793$ & $0.753$ \\
$5.0$ &     & $1.466$ & $1.429$ &     & $1.110$ & $1.055$ &     & $1.024$ & $0.974$ &     & $0.966$ & $0.913$ \\
$7.5$ &     & $2.006$ & $1.985$ &     & $1.562$ & $1.466$ &     & $1.451$ & $1.352$ &     & $1.370$ & $1.281$ \\
$10.0$&     & $2.491$ & $2.476$ &     & $1.984$ & $1.867$ &     & $1.617$ & $1.730$ &     & $1.740$ & $1.617$ \\
$15.0$&     & $3.556$ & $3.412$ &     & $2.775$ & $2.616$ &     & $2.583$ & $2.419$ &     & $2.442$ & $2.266$ \\
\hline
\end{tabular}
\caption{Numerical values of critical negative bare quadratic coupling $-m^{2}_{B,c}$ computed in the coupling basis~($-m^{2\text{ tr}}_{B,c}$) and from directly integrating the LPA potential ($-m^{2\text{ di}}_{B,c}$) for $\z= 3/4$.}
\label{tab:critlinegauss}
\end{table}

\begin{figure}[H]
    \centering
    % Left: q=3
    \begin{subfigure}[t]{0.48\textwidth}
        \centering
        \includegraphics[width=\linewidth]{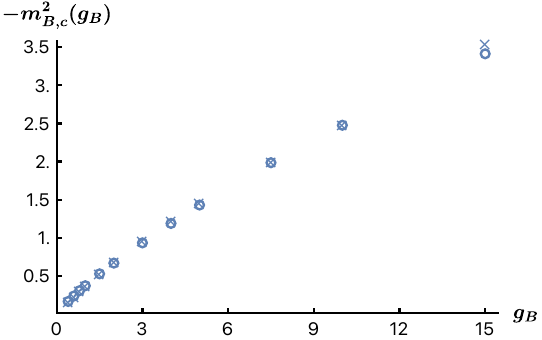}
    \end{subfigure}
    \hfill
    % Right: q=5
    \begin{subfigure}[t]{0.48\textwidth}
        \centering
        \includegraphics[width=\linewidth]{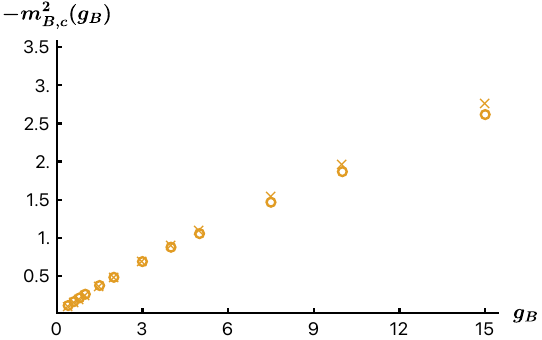}
    \end{subfigure}
    
    \caption{The critical line for $\T_3$ (left) and $\T_5$ (right) in blue and orange, respectively, for $\z = 3/4$. The circles correspond to critical points obtained via the truncation method, and the crosses correspond to critical points in the full potential formalism. }
    \label{fig:CriticalLinegauss1}
\end{figure}

\begin{figure}[H]
    \centering
    % Left: q=6
    \begin{subfigure}[t]{0.48\textwidth}
        \centering
        \includegraphics[width=\linewidth]{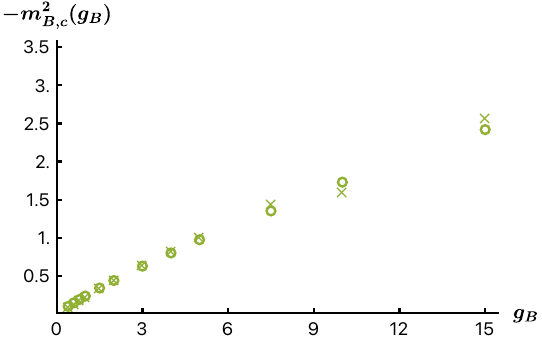}
    \end{subfigure}
    \hfill
    % Right: q=7
    \begin{subfigure}[t]{0.48\textwidth}
        \centering
        \includegraphics[width=\linewidth]{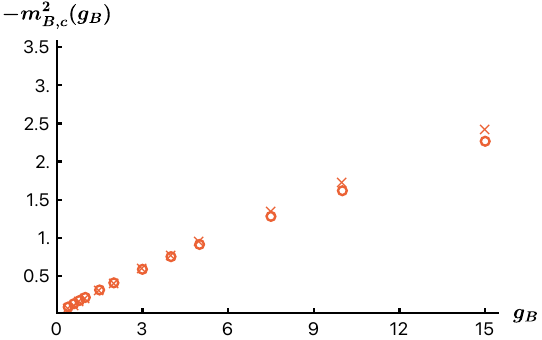}
    \end{subfigure}

    \caption{The critical line for $\T_6$ (left) and $\T_7$ (right) in green and red, respectively, for $\z = 3/4$. The circles correspond to critical points obtained via the truncation method, and the crosses correspond to critical points in the full potential formalism.}
    \label{fig:CriticalLinegauss2}
\end{figure}

Similar to the short-range theory, we plot the critical line of the Gaussian limit of long-range theory in Fig. \ref{fig:CriticalLinegauss1} and \ref{fig:CriticalLinegauss2} based on Table \ref{tab:critlinegauss}. We note that, similarly to the short-range theory, for smaller values of $g_B$, there is excellent agreement between both methods for all the values of $q$ considered. Similarly, it would be interesting to conduct a more detailed numerical investigation for a larger  range of values of $g_B$ and $q$. 

%%%%%%%%%%%%%%%%%%%%%%%%%%%%%
\section{Perturbation Theory}
\label{sec:pertRG1}
%%%%%%%%%%%%%%%%%%%%%%%%%%%%%
In addition to the FRG analysis, we examine the scalar field theory via lattice perturbation theory up to one-loop. We recall the bare action in graph units
 \begin{equation} \label{eq:BLgraphaction2}
  S[\chi]= \frac{1}{2}  \chi\cdot\left(\Delta + \mu^2\right)^\zeta\cdot\chi+ 
  \sum_i\left(\frac{1}{2} m_B^2 \chi(i)^2+\frac{g_B}{4!} \chi(i)^4\right)\,.
\end{equation}
Here, the scale $\mu$ is an explicit IR cut-off. The IR cut-off $\mu$ renders all the loop integrals convergent. For long-range models, it is technically easier to set the perturbation theory around the massless theory as in \cite{Benedetti:2020rrq}. We consider both $m^2_B$ and $g_B$ as small parameters and in a double expansion in $m^2_B$ and $g_B$, we restrict to the second order.
The one-loop effective action in graph units is given by
\begin{equation}\label{eq:1l}
 \Gamma[\phi] = S[\phi] +  \Gamma_{\LP}[\phi]\,,
\end{equation}
where $\Gamma_{\LP}$ is the one-loop correction to the effective action given by
\begin{equation} \label{eq:1loop}
   \Gamma_{\LP}[\phi] = \frac{1}{2} \Tr \log \left(\frac{S^{(2)}[\phi]}{\NN} \right)\,.
\end{equation}
Here the normalization factor has been chosen as $\NN = S^{(2)}[0]|_{m_B=0}$. At one loop, it is sufficient to consider homogeneous field configurations $\phi_0$ as there is no wave function renormalization. 
In such field configurations, the effective action $\G[\phi_0]$ reduces to just the potential part of the effective action $U(\phi_0)$, omitting the volume term. From \eqref{eq:1l} for homogeneous field configuration, we have the effective potential $U(\phi_0)$
\begin{equation} \label{eq:effpotential}
  U(\phi_0) = U_{B}(\phi_0) + U_{\LP}(\phi_0)\,.
\end{equation}
 In graph units, the one-loop correction to the effective potential, $U_{\LP}(\phi_0)$ is given by
\begin{equation} \label{eq:1lpt}
 U_{\LP}(\phi_0) =
\int_{\s(\ell)} \, d\ell \ell \rho(\ell^2) \log\left( 1+ \left\{ m^2_B + \frac{g_B}{2} \phi_0^{2}   \right\} (\ell^2 + \mu^2)^{-\zeta}   \right)\,,
\end{equation} 
in the spectral representation (similar to \eqref{eq:LPAansatz}). Considering a perturbative expansion up to $g_B^2$, $m_B^2$ and $m^2_B g_B$ from \eqref{eq:1lpt}, we find \footnote{The factor of $2$ is adjusted in the transition from \eqref{eq:1lpt} to \eqref{eq:1lpt2}, for a more lucid notation in the evaluation of the loop integrals in Appendix \ref{app:lattint}.} 
\begin{equation} \label{eq:1lpt2}
U_{\LP}(\phi_0) = \frac{1}{2}\left( m_{B}^{2} + \frac{g_{B}}{2} \phi_0^{2} \right) I_{1}(\mu) - \frac{1}{4}\left( m_{B}^{2} + \frac{g_{B}}{2} \phi_0^{2} \right)^{2} I_{2}(\mu)\,.
\end{equation}
The loop integrals present themselves as
 \begin{equation} \label{eq:loopint}
  I_{\alpha}(\mu) \coloneqq 2\int_{\s(\ell)}\frac{\ell\rho(\ell^2)}{\left(\ell^2+ \mu^2\right)^{\alpha\zeta}}\,d\ell\,.
\end{equation}
 We find an exact solution for the integral in terms of Appell $F_1$ functions \cite{zbMATH02708199, zwillinger2014table} as shown in Appendix \ref{app:lattint}. 
 
\begin{figure}[htbp]
    \centering
    \scalebox{0.7}{
        \begin{tikzpicture}
            \tikzset{
                dot/.style={circle, fill=black, inner sep=0pt, minimum size=4pt}
            }
     \begin{scope}[local bounding box=tadpole]
                \begin{feynman}
                    \vertex [dot] (v) at (0,0) {};
                    \vertex (o1) at (1.2, 0.7);
                    \vertex (o2) at (1.2, -0.7);
                    
                    \diagram* {
                        (v) -- [thick] (o1),
                        (v) -- [thick] (o2),
                    };
                    \draw [thick] (v) .. controls (-2.0, 1.2) and (-2.0, -1.2) .. (v);
                \end{feynman}
            \end{scope}
            \begin{scope}[xshift=5cm, local bounding box=bubble]
                \begin{feynman}
                    \vertex [dot] (l) at (-0.8, 0) {};
                    \vertex [dot] (r) at (0.8, 0) {};
                    
                    \vertex (i1) at (-2.0, 0.7);
                    \vertex (i2) at (-2.0, -0.7);
                    \vertex (o1) at (2.0, 0.7);
                    \vertex (o2) at (2.0, -0.7);

                    \diagram* {
                        (i1) -- [thick] (l) -- [thick] (i2),
                        (l) -- [thick, bend left=75] (r),
                        (r) -- [thick, bend left=75] (l),
                        (o1) -- [thick] (r) -- [thick] (o2),
                    };
                \end{feynman}
            \end{scope}
            \begin{scope}[xshift=10cm, local bounding box=cross]
                \begin{feynman}
                    \vertex (l) at (-0.8, 0);
                    \vertex [dot] (r) at (0.8, 0) {};
                    
                    \vertex (i) at (-2.2, 0); 
                    \vertex (o1) at (2.0, 0.7);
                    \vertex (o2) at (2.0, -0.7);

                    \diagram* {
                        (i) -- [thick] (l),
                        (l) -- [thick, bend left=75] (r),
                        (r) -- [thick, bend left=75] (l),
                        (o1) -- [thick] (r) -- [thick] (o2),
                    };
                    \draw[thick] ([xshift=-3.5pt, yshift=-3.5pt]l.center) -- ([xshift=3.5pt, yshift=3.5pt]l.center);
                    \draw[thick] ([xshift=-3.5pt, yshift=3.5pt]l.center) -- ([xshift=3.5pt, yshift=-3.5pt]l.center);
                \end{feynman}
            \end{scope}
        \end{tikzpicture}
    } 
    
    \caption{Feynman diagram representation of one-loop corrections to the two-point and four-point vertex upto second order in $m^2_B$ and $g_B$.~The left diagram represents the tadpole term $g_B I_1(\mu)$, the central diagram represents the term $g_B^2 I_2(\mu)$, and the right diagram corresponds to the term $m^2_B g_B I_2(\mu)$ with the cross denoting the $m^2_B$ vertex.}
    \label{fig:FDiagrams}
\end{figure}
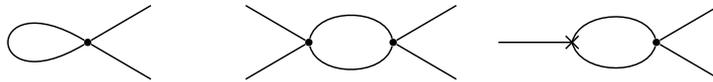

We present the loop diagrams relevant to our discussion in Fig.~\ref{fig:FDiagrams}.
 
 From \eqref{eq:effpotential} and \eqref{eq:1lpt2}, we get the four-point vertex
 \begin{equation} \label{eq:4ptbare}
  U^{(4)}(\phi_0) = g_B -\frac{3}{2}g^2_B I_2(\mu)\,,
\end{equation}
and the two-point vertex
 \begin{equation} \label{eq:2ptbare}
  U^{(2)}(\phi_0) = \mu^{2\zeta} + m^2_B + g_B I_1(\mu) -\frac{1}{2} m^2_B g_B I_2(\mu)\,.
\end{equation}

%%%%%%%%%%%%%%%%%%%%%%%%%%%%%
\paragraph{\underline{Gapless case}} \mbox{}\\
%%%%%%%%%%%%%%%%%%%%%%%%%%%%%

For the gapless theory, from \eqref{eq:loopint} (for $\alpha =2$), we find in the $\mu \to 0$ limit that the four-point function is logarithmically divergent if $\zeta = 3/4$ by virtue of the spectral dimension, $d_s =3$, as shown in Appendix \ref{app:lattint}. Therefore, we set up an $\epsilon$ expansion by tuning $\zeta= (3+\epsilon)/4$, as we would have done for long-range models in continuum three dimensions \cite{Benedetti:2020rrq}. To define RG flows, we need to introduce units where the operators and couplings of the theory are measured in terms of the scale $\mu$. Thus, based on the  spectral dimension, we define rescaled couplings $\bar g_B$ and $\bar m^2_B$ analogous to dimensionless couplings in long-range, $3$D continuum field theory
\begin{equation} \label{eq:gbarmbar}
g_B = \mu^{\epsilon} \bar g_B, \quad  m^2_B = \mu^{\frac{3+\epsilon}{2}} \bar m^2_B\,,
\end{equation}
which will lead to autonomous flow equations. 

In terms of the rescaled couplings, the four-point vertex \eqref{eq:4ptbare} is given by
\begin{equation} \label{eq:4ptbare2}
  U^{(4)}(\phi_0) = \mu^{\epsilon}\bar g_B -\frac{3}{2}\mu^{2\epsilon}\bar g^2_B I_2(\mu)\,.
\end{equation}

In the gapless theory, at small $\mu$, we have the following asymptotic behavior of the loop integral $I_2(\mu)$
\begin{equation} \label{eq:I2}
   I_{2}(\mu) \sim C_{q} \frac{\,\Gamma\left(\frac{\epsilon}{2}\right)}
{\Gamma\left(\frac{3+\epsilon}{2}\right)} \mu^{-\epsilon}\,,
\end{equation}
(see \eqref{eq:I2asympfinal} in Appendix~\ref{app:lattint}), where $C_q$ is a $q$-dependent constant.
This matches the 3D continuum result up to an overall multiplicative constant \cite{Benedetti:2020rrq}. Thus, the four-point function has the form
\begin{equation} \label{eq:4ptcl1}
     U^{(4)}(\phi) = \mu^{\epsilon}\left(\bar g_B - \frac{3}{2} \bar g^2_B \left( C_{q} \frac{\Gamma\left(\frac{\epsilon}{2}\right)}
{\Gamma\left(\frac{3+\epsilon}{2}\right)} \right)\right)\,.
\end{equation}
At leading order in small $\epsilon$, we obtain
\begin{equation} \label{eq:4ptcl}
     U^{(4)}(\phi) = \mu^{\epsilon}\left(\bar g_B - \frac{3}{2} \bar g^2_B \left(\frac{N_q}{\epsilon} + \text{const.}\right)\right)\,, \quad N_q \coloneqq 2 C_{q}.
\end{equation}
We defined $N_q \coloneqq 2 C_q$ to ensure the beta functions resemble their continuum counterparts as in \cite{Benedetti:2020rrq}. To get rid of the $\frac{1}{\epsilon}$ pole, we redefine the bare coupling in terms of a renormalized coupling $\bar g$
\begin{equation} \label{eq:rengbar}
  \bar g_B = \bar g + \frac{3}{2} \frac{N_q}{\epsilon} \bar g^2\,.
\end{equation}
This is a graph adaptation of the minimal subtraction (MS) scheme used in continuum QFT. We fix the bare coupling in graph units and adjust the renormalized coupling $\bar g$ as we tune the cut-off $\mu$ to obtain its flow.
In order to return to graph units, we multiply \eqref{eq:rengbar} with $\mu^{\epsilon}$ and get 
\begin{equation} \label{eq:barecoupl}
  g_B = \mu^{\epsilon}\left(\bar g + \frac{3}{2} \frac{N_q}{\epsilon} \bar g^2\right)\,.
\end{equation}
The beta function $\beta_{\bar g} \coloneqq \mu \frac{\partial }{\partial \mu}\bar g$ is computed by taking the logarithmic derivative of \eqref{eq:barecoupl}
\begin{equation} \label{eq:betagu}
 \epsilon\left(\bar{g}+\frac{3}{2} N_q \frac{\bar{g}^2}{\epsilon}\right)+\beta_{\bar{g}} \frac{\partial}{\partial \bar{g}}\left(\bar{g}+\frac{3}{2} N_q \frac{\bar{g}^2}{\epsilon}\right) =0\,.
\end{equation}
Solving for $\beta_{\bar g}$ up to $\bar g^2$ terms, we find
\begin{equation} \label{eq:betagpert}
 \beta_{\bar{g}} = -\epsilon \bar{g} + \frac{3}{2}\bar{g}^{2} N_{q}\,.
\end{equation}

With the running of the quartic coupling in hand, we compute the running of the quadratic coupling $\bar m^2$. The two-point vertex with rescaled couplings is
\begin{equation} 
   U^{(2)}(\phi_0) = \mu^{\frac{3+\epsilon}{2}} + \mu^{\frac{3+\epsilon}{2}}\bar m^{2}_B + \mu^{\epsilon}\bar g_B I_1 (\mu) -\frac{1}{2} \bar m_B^2 \bar g_B \mu^{\frac{3+\epsilon}{2}+\epsilon}I_2(\mu)\,.
\end{equation}
The integral $I_1 (\mu)$ in the asymptotic $\mu \to 0$ limit is regular for $\epsilon \to 0$ as demonstrated in \eqref{eq:I1asymp} of Appendix \ref{app:lattint}. The only divergent part is the $\epsilon$ pole of $I_2(\mu)$, which is absorbed by defining a renormalized mass $\bar m^2$
\begin{equation} \label{eq:renmbar}
  \bar m^2_B = \bar m^2\left(1 + \bar g \frac{N_q}{2\epsilon}\right)\,,
\end{equation}
and therefore the bare mass is 
\begin{equation} \label{eq:baremass}
   m^2_B = \mu^{\frac{3+\epsilon}{2}}\bar m^2\left(1 + \bar g \frac{N_q}{2\epsilon}\right)\,.
\end{equation}
Taking the logarithmic derivative, we obtain
\begin{equation}
 	\frac{3+\epsilon}{2} \bar m^2_B + \beta_{\bar g} \frac{\partial \bar m^{2}_{B}}{\partial \bar g}
 	+ \beta_{\bar m} \frac{ \partial \bar m^{2}_{B} }{ \partial \bar m^{2}} = 0\,.
\end{equation}
Keeping only the terms linear in $\bar g$, we get 
\begin{equation} \label{eq:betampert}
  \beta_{\bar{m}} = \frac{\bar{g} \bar{m}^2 N_{q}}{2}-\frac{3+\epsilon}{2}\bar{m}^2\,.
\end{equation}

From the beta functions, we obtain an interacting fixed-point with the critical exponents
$\left\{ -\frac{3}{2} - \frac{\epsilon}{6},\epsilon \right\}$ having one irrelevant and one relevant direction. This is consistent with the critical exponents obtained via FRG in the appropriate limits and matches the continuum result presented in \cite{Benedetti:2020rrq}. 

 We further comment that the extrapolation of the existence of an interacting fixed point in the long-range to the Wilson-Fisher fixed point in the short-range is non-trivial. In the continuum, it has been shown that at some $\z = \z^{*} < 1$, the long-range theory exhibits a crossover to the short-range theory via perturbation theory \cite{Sakcrossover73}, the FRG \cite{Solfanelli:2024obb, Defenu:2014bea}, and the conformal bootstrap \cite{Behan:2017emf, Rong:2024vxo}. Establishing such a transition on the Bethe lattice is an interesting research direction and would require analysis at higher-loop orders in perturbation theory and beyond the LPA for the FRG, which we defer to a future work.
%%%%%%%%%%%%%%%%%%%%%%%%%%%%%
\paragraph{\underline{Gapped case}} \mbox{}\\
%%%%%%%%%%%%%%%%%%%%%%%%%%%%%

For the gapped Laplacian, the loop integral in \eqref{eq:loopint} is given by
\begin{equation} \label{eq:loopintg}
  I_{\alpha}(\mu) \coloneqq 2\int^{\Lambda_\g}_{\gamma} \ell\rho_{\g}(\ell)\, d\ell\,
  \frac{1}{\left(\ell^2+ \mu^2\right)^{\alpha\zeta}}\,,
\end{equation}
which is well defined as $\mu \to 0$, for all $\a$ and $\z$. In the said limit, \eqref{eq:loopintg} evaluates to a $\a$,$\z$ and a $q$ dependent function $K_{\a,\z}(q)$ as shown in \eqref{eq:gapped}. 

Let us consider at some scale $\mu$, the renormalized coupling (in graph units) $g(\mu)$ be given by evaluating the four point vertex $U^{(4)}(\phi_0)$ at that scale. We find
\begin{equation} \label{eq:renggapped}
  g(\mu) = g_B -\frac{3}{2}g^2_B I_2(\mu)\,.
\end{equation}
Inverting the relation between bare and renormalized coupling, we get
\begin{equation} \label{eq:renggappedbare}
  g_B = g(\mu) + \frac{3}{2}g^2(\mu) I_2(\mu)\,.
\end{equation}
To derive the beta functions, we chose a scaling ansatz 
\begin{equation} \label{eq:bareggapped}
  g(\mu) = \mu^{\omega} \gb\,,
\end{equation}
where $\omega$ is some arbitrary scaling. Unlike the gapless case, we do not encounter any logarithmic divergence; therefore, a natural notion of scaling does not emerge. In scaling units, we find
\begin{equation} \label{eq:renggappedbarescale}
  g_B = \mu^{\omega}\gb + \frac{3}{2}\mu^{2\omega}\gb^2 I_2(\mu)\,.
\end{equation}
The above equation \eqref{eq:renggappedbarescale}  does not have a scaling form for general $\mu$. Nevertheless, for $\mu \to 0$, and at leading order in $\mu$, we get,
\begin{equation}
g_B =
\begin{cases}
\mu^{\omega}\,\gb, 
& \omega > 0\, \\
\displaystyle \frac{3}{2}\,K_{2,\z}(q)\,\mu^{2\omega}\,\gb^2, 
& \omega < 0\,
\end{cases}.
\label{eq:gscaling}
\end{equation}
Both of the cases lead to a tree-level flow equation,
\begin{equation} \label{eq:gappedflowg}
  \b_{\gb} = -\omega g\,,
\end{equation}
which only leads to a Gaussian fixed point. For $\omega=0$, the beta function at the scaling limit identically vanishes. 

Consider that at some scale $\mu$ the reference two-point vertex is given by
$ U^{(2)}(\phi_0)= m^2(\mu) + \mu^{2\z}
  + \left(g(\mu) + \frac{3}{2} g^{2}(\mu)\right) I_1(\mu)$. Thus, the renormalized quadratic coupling (in graph units) satisfies
\begin{equation} \label{eq:renmgapped}
   m^{2}(\mu) = m^2_B - \frac{1}{2} m^2_B g_B I_2(\mu)\,.
\end{equation}
Using \eqref{eq:renggapped} and \eqref{eq:renmgapped}, we can express $m^2_B$ as
\begin{equation} \label{eq:renmgappedbare}
  m^{2}_B = m^2(\mu) + \frac{1}{2}g(\mu) m^2(\mu)I_2 (\mu) \,,
\end{equation}
By virtue of being a quadratic coupling, we impose the scaling
\begin{equation} \label{eq:baremgapped}
  m^{2}(\mu) = \mu^{2\z} \mb\,,
\end{equation}
where, by definition of long-range and short-range models $0<\z\leq1$.
Thus in scaling units,
\begin{equation} \label{eq:}
  m^{2}_B  =
  \mu^{2\z}\mb^2
  + \frac{1}{2} \mu^{2\z+\omega}I_2(\mu) \gb \mb^2\,,
\end{equation}
which again, like the four-point vertex, does not have a scaling form for general $\mu$. Nevertheless, for vanishing $\mu$, we do have the following scaling relations,
\begin{equation}
m_B^2 =
\begin{cases}
\displaystyle \frac{1}{2}\,\mu^{2\z + \omega}\,K_{2,\z}(q)\,\gb\,\mb^2
& \omega < 0\, \\
\mu^{2\z}\,\mb^2
& \omega > 0\,
\end{cases}.
\label{eq:mscaling}
\end{equation}
Similar to the quartic coupling, both cases lead to the same tree-level flow equation
\begin{equation} \label{eq:mflowgapped}
  \b_{\mb} = -2\z \mb^2\,,
\end{equation}
which ultimately only admits a Gaussian fixed point.
\section{Results and Discussion}
\label{sec:results}
In this paper, we studied the critical properties of a $\Z_2$ symmetric one-component scalar field theory on the Bethe lattice, via two complementary methods, namely lattice FRG and lattice perturbation theory. We present a summary of our results on critical exponents on the Bethe lattice in Table \ref{tab:results_summary}.

\begin{table}[H]
    \centering
        \renewcommand{\arraystretch}{2.2} 
    \setlength{\tabcolsep}{4pt} 

    \begin{tabularx}{\textwidth}{>{\centering\arraybackslash}m{2.4cm} >{\centering\arraybackslash}m{2.4cm} |Y|Y|Y|}
        \cline{3-5}
        \multicolumn{1}{c}{} & \multicolumn{1}{c|}{} & $\z=1$ \par (short-range) & $\z= 3/4$ \par (long-range at marginality) & $\z = (3+\epsilon)/4$ \par (small-$\epsilon$ expansion) \\ \hline

        \multicolumn{1}{|c|}{\multirow{3}{*}{FRG (LPA)}} 
        & \multicolumn{1}{c|}{IR fixed point} & Wilson-Fisher & Gaussian & interacting \\ \cline{2-5}
        \multicolumn{1}{|c|}{} & \multicolumn{1}{c|}{truncation order} & $\OO(16)$ & $\OO(16)$ & $\OO(4)$ \\ \cline{2-5}
        \multicolumn{1}{|c|}{} & \multicolumn{1}{c|}{critical exponent} & $\th_{1,1} = -1.541$ & $\th_{1,3/4} = -3/2$ & $\th_{1,(3+\epsilon)/4} = - 3/2 - \epsilon/6$ \\ \hline
     
        \multicolumn{1}{|c|}{\multirow{2}{*}{\makecell{perturbation \\ theory (1-loop)}}} 
    
        & \multicolumn{1}{c|}{IR fixed point \vphantom{$\th_{1,(3+\epsilon)/4} = - 3/2 - \epsilon/6$}} & - & - & interacting \\ \cline{2-5}
        \multicolumn{1}{|c|}{} & \multicolumn{1}{c|}{critical exponent} & - & - & $\th_{1,(3+\epsilon)/4} = - 3/2 - \epsilon/6$ \\ \hline
    \end{tabularx}
    
    \caption{Summary of results of critical exponents. The critical exponents of $\mathbb{Z}_2$-symmetric scalar field theory with a kinetic term of the form $\Delta_0^{\z}$ (gapless Laplacian) are derived from the beta functions up to $\mathcal{O}(16)$ in the FRG framework within the LPA, as discussed in Section~\ref{sec:cbasis}. Particular emphasis is placed on $\z=1$, corresponding to the standard short-range case; $\z=3/4$, where the stable fixed point is Gaussian; and a small-$\epsilon$ expansion with $\z=(3+\epsilon)/4$, allowing for comparison with one-loop perturbation theory.
In addition, one-loop beta functions and critical exponents are obtained using perturbation theory for the same small-$\epsilon$ expansion. The results are consistent with the FRG analysis at $\mathcal{O}(4)$ and with continuum perturbation theory.}
    \label{tab:results_summary}
\end{table}
In Section \ref{sec:funRG}, we considered both short- and long-range models in the framework of FRG and obtained the flow equations in the LPA. In Section \ref{sec:cbasis}, we found that, in the FRG framework, tuning the Laplacian to a gapless regime allowed access to non-trivial fixed points. In the scaling limit, the flow equations resemble those of continuum three-dimensional flat space, because the spectral dimension is three. 
For the computation of critical exponents, we focused on two particular cases: first, the short-range theory ($\z=1$), which flows to the Wilson-Fisher fixed point, and second, when $\z=3/4$, where the theory becomes marginal and flows to the Gaussian fixed point. We also obtained the critical line of the theory in the hopping parameterization of the bare action for $\z=1$ and in terms of bare couplings for $\z = 3/4$ by numerically solving the beta functions up to an $\mathcal{O}(16)$ truncation in Mathematica. Complementarily, we also computed the critical line by directly integrating the LPA flow equation~\eqref{eq:optimizedLPAflow}. The two methods agree well with each other for small values of the degree $q$ and coupling constant $\l$, as shown in Tables \ref{tab:critline} and \ref{tab:critlinegauss} and Figures \ref{fig:CriticalLine12}, \ref{fig:CriticalLinegauss1}, and \ref{fig:CriticalLinegauss2}.

Similarly, in perturbation theory discussed in Section \ref{sec:pertRG1}, the gapless Laplacian was required to find an interacting fixed point. In the gapped case, the loop integrals did not exhibit any divergences for any $\z$, hence only the tree-level beta functions were found exhibiting a Gaussian fixed point. On the contrary, for the gapless theory, at $\z= 3/4$, the four-point function exhibited logarithmic divergences at one-loop. This allowed us to set up an $\epsilon$-expansion by considering $\z= (3+\epsilon)/4$ with $\epsilon>0$ to tune away from marginality, and find an interacting fixed point. The corresponding critical exponents agree with the continuum results found in \cite{Benedetti:2020rrq}, and with the results derived from FRGE within the LPA at an $\mathcal{O}(4)$ truncation for small $\epsilon$.

The existence of the Wilson-Fisher fixed point for the short-range scalar field theory on the Bethe lattice is contrary to the expectation from the Ising model, which also has $\Z_2$ symmetry, but exhibits spontaneous magnetization with mean-field critical exponents. A few comments are in order; for the short-range theory, tuning to the gapless regime is achieved by adding a negative local quadratic term proportional to the gap. Such a deformation is not possible in the Ising Hamiltonian, as any term of type $\s(i)^{2n}$ will be a constant as the Ising spins $\s$ take values of $\pm 1$. It seems that, in spite of the same internal symmetries, the use of a continuous degree of freedom in the scalar field theory, instead of a discrete spin variable, changes the universality class. This raises an interesting question: What characterizes universality classes in hyperbolic spaces, beyond internal symmetries alone?

The key finding of our paper is the indication of the existence of non-interacting fixed points for scalar field theories with $\Z_2$ symmetries, which opens up numerous interesting research avenues.  One direction would be to go beyond the LPA and one-loop analysis, and consider the flow of derivative terms. This would provide a deeper understanding of the critical structure of scalar field theories on the Bethe lattice. In conjunction with this, it would enable a systematic investigation of the long-range to short-range crossover, and a search for further evidence for the existence of a Wilson–Fisher fixed point on the Bethe lattice. Another natural extension of this work would be to generalize the present analysis to general regular hyperbolic lattices and to models with other internal symmetries (such as $O(N)$), as well as fermionic fields, to investigate whether such an interacting fixed point and non-mean-field behavior exist. 
Another worthwhile avenue is to explore the intermediate phase suggested in field theoretic studies \cite{Doyon:2004fv, Carmi:2018qzm} and lattice models \cite{Breuckmann_2020, Rietman1992TheIM,wu2000ising, Okunishi_2024}, by looking at spatially dependent vacuum configurations. This could potentially enhance our understanding of the nature of phase transitions on hyperbolic space and associated universality classes.

\section*{Acknowledgements}

S.S. would like to thank Debasish Banerjee, Dario Benedetti, Kedar Damle, Nadia Flodgren, Razvan Gurau, Michael Lathwood, Cihan Pazarbasi, Shinobu Hikami, Philipp Hoehn, and Yasha Neiman for fruitful and enlightening discussions. S.S would also like to thank the Insitute Henri Poincar\'e thematic program ``Random Tensors", 30 September - 18 October 2024, and ``42nd International Symposium on Lattice Field Theory" at Tata Institute of Fundamental Research (TIFR) from November 2-8, 2025, where part of the discussions took place.

\newpage
\appendix

%%%%%%%%%%%%%%%%%%%%%%%%%%%%%
\section{Large $q$ limit of Ultra local initial conditions}
\label{app:ULC}
%%%%%%%%%%%%%%%%%%%%%%%%%%%%%

In this appendix, we will demonstrate that the ultra-local initial conditions for large $q$ is the Gaussian fixed point. Let us recall from \eqref{eq:singsite} that, at the ultra-local scale $k = \Lambda_0$ the functional integro-differential equation for the scale dependent effective potential is given by
\begin{equation} \label{eq:singsiteA}
e^{-\hat{U}_{\Lambda_0}(\phi_0)+ \phi_0 \,\hat{U}_{\L_0}^{(1)}(\phi_{0})}
=\int^{\infty}_{-\infty} d \chi\,
e^{-\frac{1}{2}\left(m_B^2+\L_0^{2\zeta }\right) \chi^2
-\frac{g_{B}}{4!} \chi^4
+\chi\,\hat{U}_{\L_0}^{(1)}(\phi_{0}) }\,.
\end{equation}

We use transformation rules \eqref{eq:auttransform} adapated to the ultra-local scale
\begin{equation} \label{eq:auttransformA}
  \hat{V}_{\L_0}(\vph) \coloneqq \frac{1}{c_q \L_0^{d_s}}\hat{U}_{\L_0}(\phi_0(\vph))\,,\quad 
 \phi_0(\vph)\coloneqq \L_0^{\frac{d_s-2\z}{2}}\sqrt{c_q}\,\vph\,,
\end{equation}
and plug in the coupling basis \eqref{eq:tansatz} into \eqref{eq:singsiteA} to obtain the initial conditions $g_{2i}(\t=1)$ (or equivalently $g_{2i}(k= \L_0)$).

In the transformed variables, \eqref{eq:singsiteA} reads as
\begin{equation}\label{eq:dimlessULCA}
	e^{-c_q \L_0^{d_s}\left( \hat{V}_{\L_0}(\vph)- \vph \hat{V}_{\L_0}^{(1)}(\vph) \right)}
 = \int d \chi\,
 e^{-\frac{1}{2}\left(m_B^2+\L_0^{2\zeta }\right) \chi^2
 -\frac{g_{B}}{4!} \chi^4
 +(c_q \L_0^{d_s})(\L_0^{\frac{2\z-d_s}{2}}c_q^{-1/2})\,\chi\,\hat{V}_{\L_0}^{(1)}(\vph) }\,.
\end{equation}

From the definitions of $\L_0$ and $c_q$, we find the following large $q$ asymptotics
\begin{equation} \label{eq:asympq}
c_q \L_0^{d_s} \sim \frac{16 \z}{3 \pi }\,, \quad
\frac{\L_0^{\frac{2\z-d_s}{2}}}{\sqrt{c_q}} \sim
\frac{\sqrt{3 \pi } 2^{\zeta -2} q^{\zeta /4}}{\sqrt{\zeta }}\,, \quad
\L_0^{2\zeta} \sim  (16 q)^{\zeta /2}.
\end{equation}

We perform a change of variables $\chi^{\prime} = A(q)\chi$, where
$A(q) \coloneqq \frac{\L_0^{\frac{2\z-d_s}{2}}}{\sqrt{c_q}}$, and we get
\begin{equation}\label{eq:dimlessULCA2}
	e^{-c_q \L_0^{d_s}\left( \hat{V}_{\L_0}(\vph)- \vph \hat{V}_{\L_0}^{(1)}(\vph) \right)}
 = A(q)^{-1} \int d \chi^{\prime}\,
 e^{-\frac{1}{2}\left(m_B^2+\L_0^{2\zeta }\right)
  \left(\frac{\chi^{\prime}}{A(q)}\right)^2
 -\frac{g_{B}}{4!} \left( \frac{\chi^{\prime}}{A(q)}\right)^4
 +(c_q \L_0^{d_s})\,\chi^{\prime}\,\hat{V}_{\L_0}^{(1)}(\vph) }\,.
\end{equation}

 In the $q \to \infty$ regime, we observe that $A(q)$ grows as $q^{\z/4}$. Thus the quartic term is suppressed and the Gaussian term dominates and we get using \eqref{eq:asympq}
\begin{equation}\label{eq:dimlessULCGaussA}
	e^{- \frac{16 \z}{3\pi}\left( \hat{V}_{\L_0}(\vph)- \vph \hat{V}_{\L_0}^{(1)}(\vph) \right)}
 = A(q)^{-1}\int d \chi^{\prime}\,
 e^{-\frac{1}{2}\L_0^{2\z} \left(\frac{\chi^{\prime}}{A(q)}\right)^{2}
 +\, \frac{16 \z}{3 \pi }\chi^{\prime}\,\hat{V}_{\L_0}^{(1)}(\vph) }\,.
\end{equation}

Here we have not explicitly written the value of $A(q)$ and $\L_0$ for brevity. The Gaussian integration leads to the following differential equation for $\hat{V}_{\L_0}(\vph)$
\begin{equation} \label{eq:diffV}
   \frac{16 \z}{3\pi}\left( -\hat{V}_{\L_0}(\vph)+ \vph \hat{V}_{\L_0}^{(1)}(\vph) \right)
=\frac{ 128\,\z^2 A^2(q)\hat{V}_{\Lambda_0 }^{(1)}(\varphi ){}^2}
{ 9 \pi^2 \Lambda_0 ^{2\zeta }}
+\frac{1}{2} \log \left(\frac{2 \pi A^2(q) }{\Lambda_0 ^{2\zeta }}\right)\,.
\end{equation}

To solve \eqref{eq:diffV}, we differentiate with respect to $\vph$ and obtain the following auxillary differential equation
\begin{equation} \label{eq:intermediatestep}
  \hat{V}^{(2)}(\vph)
  \left\{ \frac{16 \z \vph}{3\pi}
  - \left(\frac{ 8 \z A(q)}{ 3 \pi\L_0^{\z}}\right)^2
  \hat{V}^{(1)}(\vph) \right\} =0\,,
\end{equation}
which leads to two linear ODEs. $\hat{V}_{\L_0}^{(2)}(\vph) = 0$ is an unphysical solution, as it refers to the limit of the two-point vertex not existing. The other solution corresponds to
\begin{equation} \label{eq:GaussSolution}
  \hat{V}_{\L_0}(\vph)
  = \frac{3 \pi  \Lambda _0^{2 \zeta }}{8 \zeta  A(q)^2} \vph^2
  + \text{const.}
  = \frac{1}{2} \vph^2 + \text{const.}\,,
\end{equation}
after substituting large $q$ values from \eqref{eq:asympq}, and without loss of generality we can choose the integration constant to be zero.

As a consequence of \eqref{eq:auttransformA}, we have
$\hat{V}_{\L_0}(\vph) = V_{\L_0}(\vph) + \frac{1}{2} \vph^2$,
therefore $V_{\L_0}(\vph) = 0$. In the truncation basis \eqref{eq:tansatz}, the ultra-local initial condition from \eqref{eq:GaussSolution} is simply
$g_{2i}(k= \L_0) = 0$ for all $i \geq 1$. Thus, the large-$q$ limit is governed by the Gaussian fixed point.

%%%%%%%%%%%%%%%%%%%%%%%%%%%%%
\section{Lattice loop Integrals}
\label{app:lattint}
%%%%%%%%%%%%%%%%%%%%%%%%%%%%

\textbf{Gapless Laplcian:} For the gapless Laplacian $\D_{0 }$, recall that the density of states $\rho_0(\ell^2)$ \eqref{eq:Gaplless}
is
\begin{equation} \label{eq:dos2a}
 \rho_0(\ell^2) = \frac{q}{2\pi} \frac{\sqrt{\ell ^2}\sqrt{\L_0^2-\ell^2} }{(\ell^2+\g^2)(\L^2_\g-\ell^2)}\,.
\end{equation}
In the proceeding computation of the loop integral, it will be convenient to perform the change of variables $ \ell^2 = z $. Therefore, we get
\begin{equation} \label{eq:dos2}
 \rho_0(z) = \frac{q}{2\pi} \frac{\sqrt{z} \sqrt{\L_0^2-z} }{(z+\g^2)(\L^2_\g-z)}\,.
\end{equation}
We perform a partial fraction decomposition of $\rho_0(z)$
\begin{equation} \label{eq:Dospartfrac}
  \rho_0(z) = \frac{1}{4\pi}  \left\{\frac{\sqrt{\L_0^2-z} \sqrt{z}}{(\L^2_\g-z)} +  \frac{\sqrt{\L_0^2-z} \sqrt{z}}{(\g^2+z)}\right\}\,,
\end{equation}
where $\frac{q}{2\pi}\frac{1}{\L^2_\g + \g^2} = \frac{1}{4\pi}$, and we define $\rho_{0,1}(z) \coloneqq  \frac{1}{4\pi}\frac{\sqrt{\L_0^2-z} \sqrt{z}}{(\L^2_\g-z)}$ and $\rho_{0,2}(z)\coloneqq \frac{1}{4\pi}\frac{\sqrt{\L_0^2-z} \sqrt{z}}{(\g^2+z)}$.
\newline

The loop integral that we are interested in is  
 \begin{equation} \label{eq:loopint2ap}
  I_{\a}(\mu) \coloneqq \int_{0}^{\L_0^2} \rho_0(z) \frac{1}{\left(z+ \mu^2\right)^{\a\z}}\, dz\,,
\end{equation}
and by using the partial fraction decomposition, we split the integral into two,
\begin{equation} \label{eq:looopintsplit}
    I_{\a,i}(\mu) \coloneqq \int_{0}^{\L_0^2} \rho_{0,i}(z) \frac{1}{\left(z+ \mu^2\right)^{\a\z}}\, dz\,,
    \quad
    i = 1, \, 2\,.
\end{equation}

Let us recall the integral representation of the Appell $F_1$ function \cite{zwillinger2014table,zbMATH02708199} is given by
\begin{equation} \label{eq:AppellF1}
 F_1\left(a;b_1,b_2;c;x_1,x_2\right)=\frac{\Gamma (c) }{\Gamma (a) \Gamma (c-a)}\int_0^1 t^{a-1} (1-t)^{-a+c-1} 
 \left(1-t x_1\right)^{-b_1} \left(1-t x_2\right)^{-b_2} \, dt\,,
\end{equation}
when $\operatorname{Re}(a)>0$ and $\operatorname{Re}(c-a)>0$. The integrals $I_{\a,i}$ can be recast to resemble the integral representation of the Appell $F_1$ function. Focusing on $I_{\a,1}$, we rewrite the integral into the following form
\begin{align}
	I_{\a,1}(\mu)&= \frac{1}{4\pi}\int_0^{\L_0^2} z^{1/2} (\L_0^2-z)^{1/2}(\L^2_\g-z)^{-1}(\mu^2+z)^{-\a\z}\, dz\nonumber\\
	&= \frac{(\L_0^2)^{1/2}\mu^{-2\a\z}}{4\pi\L^2_\g}\int_0^{\L^2} z^{1/2}\left(1-\frac{z}{\L_0^2}\right)^{1/2}\left(1-\frac{z}{\L^2_\g}\right)^{-1}
        \left(1+ \frac{z}{\mu^2}\right)^{-\a\z} dz\,,
\end{align}
and then we perform a change of variable $z^{\prime} = \frac{z}{\L^2}$ to write
\begin{equation}\label{eq:Ia1}
	I_{\a,1}(\mu)=  \frac{\L_0^4 \mu^{-2\a\z}}{4\pi\L^2_\g}\int_0^{1} z^{\prime}{}^{1/2}\left(1-z^{\prime}\right)^{1/2}
        \left(1-\frac{\L_0^2z^{\prime}}{\L^2_\g}\right)^{-1}
        \left(1+ \frac{\L_0^2z^{\prime}}{\mu^2}\right)^{-\a\z} dz^{\prime}\,.
\end{equation}	
A similar exercise reveals that $I_{\a,2}$ has a similar form
\begin{equation}\label{eq:Ia2}
	I_{\a,2}(\mu)=  \frac{\L_0^4 \mu^{-2\a\z}}{4\pi\g^2}\int_0^{1} z^{\prime}{}^{1/2}\left(1-z^{\prime}\right)^{1/2}
        \left(1+\frac{\L_0^2z^{\prime}}{\g^2}\right)^{-1}
        \left(1+ \frac{\L_0^2z^{\prime}}{\mu^2}\right)^{-\a\z} dz^{\prime}\,.
\end{equation}
By comapring \eqref{eq:AppellF1} to \eqref{eq:Ia1} and \eqref{eq:Ia2}, we get,
\begin{equation} \label{eq:F1pars}
  a = \frac{3}{2}, \quad c = 3,\;\; b_1 = 1,\;\; b_2 = \a\z,\;
  x_1 = \frac{\L_0^2}{\L^2_\g}\,\text{or } -\frac{\L_0^2}{\g^2},\;\;
  x_2 = -\frac{\L_0^2}{\mu^2}.
\end{equation}
Clearly,  $\Re(a)>0$ and $\Re(c-a)>0$, therefore
\begin{equation}\label{eq:Ia1appl}
	I_{\a,1}(\mu) = \mu^{-2\a\z} \frac{\L_0^4}{32 \L^2_\g}
        F_1\left(\frac{3}{2};1,\a\z;3;\frac{\L_0^2}{\L^2_\g},-\frac{\L_0^2}{\mu^2}\right)\,,
\end{equation}
and
\begin{equation}\label{eq:Ia2appl}
	I_{\a,2}(\mu) = \mu^{-2\a\z} \frac{\L_0^4}{32 \g^2}
        F_1\left(\frac{3}{2};1,\a\z;3;-\frac{\L_0^2}{\g^2},-\frac{\L_0^2}{\mu^2}\right)\,.
\end{equation}
 In the scaling limit, we are interested in the asymptotic behavior of the loop integrals as $\mu \to 0$. We  present the asymptotic analysis case by case.

%%%%%%%%%%%%%%%%%%%%%%%%%%%%
\subparagraph{Case 1: $\a = 1$}\mbox{}\\
%%%%%%%%%%%%%%%%%%%%%%%%%%%%
 
 The  asymptotic behavior as $\mu \to 0$ can be directly obtained from the integrals \eqref{eq:Ia1} and \eqref{eq:Ia2}. In the said limit, $1+ \frac{\Lambda_0^2 z^{\prime}}{\mu^2} \sim \frac{\Lambda_0^2 z^{\prime}}{\mu^2}$ therefore

 \begin{align} 
  I_{1,1}(\mu ) &\sim \frac{\Lambda_0^{\frac{5-\epsilon }{2} }}{4\pi\Lambda^2_\g}\int_0^{1} z^{\prime}{}^{-(1+\epsilon)/4}\left(1-z^{\prime}\right)^{1/2}\left(1-\frac{\Lambda_0^2 z^{\prime}}{\Lambda^2_\g}\right)^{-1} dz^{\prime}\,, \label{eq:I11}\\
    I_{1,2}(\mu ) &\sim \frac{\Lambda_0^{\frac{5-\epsilon }{2} }}{4\pi\g^2}\int_0^{1} z^{\prime}{}^{-(1+\epsilon)/4}\left(1-z^{\prime}\right)^{1/2}\left(1+\frac{\Lambda_0^2 z^{\prime}}{\g^2}\right)^{-1} dz^{\prime}\,, \label{eq:I12}
\end{align}
where the $z^{\prime}$ integrals evaluate to hypergeometric functions\cite{zwillinger2014table}. Thus, we conclude that
\begin{equation} \label{eq:I1asymp}
   I_{1}(\mu) \sim C^{\prime}_{q}  \frac{\Gamma\left(\frac{3-\epsilon }{4}\right)}{\Gamma\left(\frac{9-\epsilon }{4}\right)}\,.
\end{equation}
Here, $C^{\prime}_{q}$ is a constant coming from the hypergeometric function, $\Gamma(3/2)$ and other $q$ dependent prefactors.
 
 Thus, $I_1(\mu)$ is independent of $\mu$ in the scaling limit and is regular at  $\epsilon =0$.
 
%%%%%%%%%%%%%%%%%%%%%%%%%%%%
\subparagraph{Case 2: $\a = 2$}\mbox{}\\
%%%%%%%%%%%%%%%%%%%%%%%%%%%%%

For $I_{2,i}(\mu)$, the integrals encounter a singularity at $z =0$ for vanishing $\mu$ and therefore requires additional treatment. Focusing on $I_{2,1}(\mu)$, we split the integral \eqref{eq:Ia1} into two domains
\begin{equation} \label{eq:I21spli}
\begin{aligned}
 I_{2,1}(\mu)&=   \frac{\Lambda_0^{4}\mu^{-3-\epsilon}}{4\pi\Lambda^2_\g}\int_0^{\delta} z^{\prime}{}^{1/2}\left(1-z^{\prime}\right)^{1/2}\left(1-\frac{\Lambda_0^2 z^{\prime}}{\Lambda^2_\g}\right)^{-1}\left(1+ \frac{\Lambda_0^2 z^{\prime}}{\mu^2}\right)^{-\frac{3+\epsilon}{2}} dz^{\prime}\,\\
 &+   \frac{\Lambda_0^{4}\mu^{-3 -\epsilon}}{4\pi\L_x\g^2}\int_{\delta}^{1} z^{\prime}{}^{1/2}\left(1-z^{\prime}\right)^{1/2}\left(1-\frac{\Lambda_0^2 z^{\prime}}{\Lambda^2_\g}\right)^{-1}\left(1+ \frac{\Lambda_0^2 z^{\prime}}{\mu^2}\right)^{-\frac{3+\epsilon}{2}} dz^{\prime}\,,
 \end{aligned}
\end{equation}
where the second term is regular and only leads to a constant contribution. For sufficiently small $\d$,
the factors $\left(1-z^{\prime}\right)^{1/2}\left(1-\frac{\Lambda_0^2 z^{\prime}}{\Lambda^2_\g}\right)^{-1} \sim 1$ and consequently, \eqref{eq:I21spli} in asymptotic $\mu \to 0$ regime is given by,
\begin{equation} \label{eq:ol}
   I_{2,1}(\mu) \sim  \frac{\Lambda_0^{4}\mu^{-3-\epsilon}}{4\pi\Lambda^2_\g} \int_0^{\delta} z^{\prime}{}^{1/2}\left(1+ \frac{\Lambda_0^2 z^{\prime}}{\mu^2}\right)^{-\frac{3+\epsilon}{2}} dz^{\prime}\,.
\end{equation}
We perform a change of variables $t = \frac{\Lambda_0^2 z^{\prime}}{\mu^2}$ and obtain
\begin{equation} \label{eq:olchangedt}
I_{2,1}(\mu) \sim 
\frac{\Lambda_0\, \mu^{-\epsilon}}{4\pi\, \Lambda^2_\g}
\int_{0}^{\frac{\Lambda_0^2\delta}{\mu^{2}}}
t^{1/2}\,(1+t)^{-\frac{3+\epsilon}{2}}\,dt\,,
\end{equation}
and for small $\mu$ can be approximated by
\begin{equation} \label{eq:olchangedtinfty}
I_{2,1}(\mu) \sim 
\frac{\Lambda_0\, \mu^{-\epsilon}}{4\pi\, \Lambda^2_\g}
\int_{0}^{\infty}
t^{1/2}\,(1+t)^{-\frac{3+\epsilon}{2}}\,dt\,.
\end{equation}
The integral \eqref{eq:olchangedtinfty} evaluates to 
 \begin{equation} \label{eq:I21asymp}
  I_{2,1}(\mu) \sim C_{q,1} \frac{\,\Gamma\left(\frac{\epsilon}{2}\right)}
{\Gamma\left(\frac{3+\epsilon}{2}\right)} \mu^{-\epsilon}
\,.
\end{equation}
A similar analysis for $I_{2,2}(\mu)$ yields
 \begin{equation} \label{eq:I22asymp}
  I_{2,2}(\mu) \sim C_{q,2} \frac{\,\Gamma\left(\frac{\epsilon}{2}\right)}
{\Gamma\left(\frac{3+\epsilon}{2}\right)} \mu^{-\epsilon}
\,.
\end{equation}
Therefore, the entire integral $I_2(\mu)$ has the following asymptotic form
 \begin{equation} \label{eq:I2asympfinal}
  I_{2}(\mu) \sim C_{q} \frac{\,\Gamma\left(\frac{\epsilon}{2}\right)}
{\Gamma\left(\frac{3+\epsilon}{2}\right)} \mu^{-\epsilon}
\,,
\end{equation}
where $C_q \coloneqq C_{q,1} + C_{q,2}$. This integral has an $\epsilon$ pole corresponding to a logarithmic divergence for $\z = 3/4$ for the four point function. We point out that the exact value of $C_q$ is insignificant for our scaling analysis in Section \ref{sec:pertRG1}.

\textbf{Gapped Laplcian:} For the gapped Laplacian, the loop integral in \eqref{eq:loopint} (in $\ell^2 =z$) is given by
\begin{equation} \label{eq:loopintgA}
  I_{\alpha}(\mu) = \int^{\Lambda^2_\g}_{\gamma^2} 
  \frac{\rho_{\g}(z)}{\left(z + \mu^2\right)^{\alpha\zeta}} dz\,.
\end{equation}
To write in terms of the integral representation of the Appell $F_1$ function, we change $z \to z + \g^2$, and we get
\begin{equation} \label{eq:loopintgA2}
  I_{\alpha}(\mu) = \int^{\Lambda^2_0}_{0} 
  \frac{\rho_{\g}(z + \g^2 )}{\left(z +\g^2 + \mu^2\right)^{\alpha\zeta}} dz = \int^{\Lambda^2_0}_{0} 
  \frac{\rho_{0}(z)}{\left(z +\g^2 + \mu^2\right)^{\alpha\zeta}} dz \,,
\end{equation}
 where we have used $\rho_{\g}(z+\g^2) = \rho_0(z)$. Essentially, this boils down to the computations of the loop integrals of the gapless Laplacian with $\mu^2 \to \mu^2 + \g^2$. We are particularly interested in the scaling limit where $\mu \to 0$. In this limit,
  \begin{equation} \label{eq:kzeroloop}
  I_{\alpha}(0) = \int^{\Lambda^2_0}_{0} z^{1/2}(\L^2_0- z)(\L^2_{\g}-z)(z+\g^2)^{-1-\a\z}\,,
\end{equation}
evaluates to a $q$, $\a$ and $\z$ dependent function in terms of Appell $F_1$ function 
\begin{equation} \label{eq:gapped} 
  I_{\alpha}(0) = K_{\a,\z}(q) = \frac{4}{15} \Lambda_0^5 \Lambda_\gamma^2\left(\gamma^2\right)^{-(1+\alpha \zeta)} F_1\left(\frac{3}{2} ;1,1+\alpha \zeta ; 3 ; \frac{\Lambda_0^2}{\Lambda_\gamma^2},-\frac{\Lambda_0^2}{\gamma^2}\right)\,.
\end{equation}

\let\oldbibliography\thebibliography 
\renewcommand{\thebibliography}[1]{\oldbibliography{#1}
\setlength{\itemsep}{-1pt}}
\bibliographystyle{JHEP}
\bibliography{references.bib}

\end{document}